\documentclass[11pt,a4paper]{article}

\usepackage{jheppub}
\usepackage[usenames,dvipsnames]{xcolor}
 
\usepackage{amssymb} 
\usepackage{amsmath}
\usepackage{mathtools}
\usepackage{amsfonts}    
\usepackage{dsfont}
\usepackage{pdfpages}
\usepackage{verbatim}
\usepackage{stmaryrd}
\usepackage{graphicx}
\usepackage{tikz}
\usetikzlibrary{arrows}
\usetikzlibrary{decorations.pathmorphing}
\usetikzlibrary{ decorations.markings}
\usetikzlibrary{shapes.misc}
\tikzset{cross/.style={cross out, draw=black, minimum size=2*(#1-\pgflinewidth), inner sep=0pt, outer sep=0pt},
cross/.default={1pt}}
\usetikzlibrary{shapes.geometric}
\tikzset{
    partial ellipse/.style args={#1:#2:#3}{
        insert path={+ (#1:#3) arc (#1:#2:#3)}
    }
}

\renewcommand{\figurename}{Fig.}

\newcommand{\bi}{\begin{itemize}}
\newcommand{\ei}{\end{itemize}}
\newcommand{\bea}{\begin{eqnarray}}
\newcommand{\eea}{\end{eqnarray}}
\newcommand{\be}{\begin{equation}}
\newcommand{\ee}{\end{equation}}

 \definecolor{pink}{rgb}{1.0, 0.13, 0.32}

\numberwithin{equation}{section}
\allowdisplaybreaks  % allow page breaks in displayed eqs

\begin{document}

\vspace*{2.5cm}
\begin{center}
{ \Large \textsc{Features of the Partition Function of a $\Lambda>0$ Universe}\\}
\vspace*{1.7cm}

\vspace*{1.7cm}

Dionysios Anninos,$^{1,2}$ Chiara Baracco,$^1$ Samuel Brian,$^1$ and Frederik Denef$^3$ \\ 

\vspace*{0.6cm}
%\newline
\vskip4mm
\begin{center}
{
\footnotesize
{$^1$Department of Mathematics, King's College London, Strand, London WC2R 2LS, UK \newline
$^2$Instituut voor Theoretische Fysica, KU Leuven, Celestijnenlaan 200D, B-3001 Leuven, Belgium \newline
$^3$Department of Physics, Columbia University, 538 West 120th Street, New York, New York 10027, U.S.  \\
}}
\end{center}

\begin{center}
{\textsf{\footnotesize{
dionysios.anninos@kcl.ac.uk, samuel.brian@kcl.ac.uk, chiara.baracco@kcl.ac.uk}, fmd7@columbia.edu} } 
\end{center}

%\vspace*{0.6cm}

%\vspace*{0.8cm}

\vspace*{1.5cm}
\begin{abstract}
\noindent

\end{abstract}
\vspace*{0.6cm}

\end{center}
%\vspace*{1.5cm}
\begin{abstract}
\noindent

\end{abstract}
We consider properties of the gravitational path integral, $\mathcal{Z}_{\text{grav}}$, of a four-dimensional gravitational effective field theory with $\Lambda>0$ at the quantum level. To leading order, $\mathcal{Z}_{\text{grav}}$ is dominated by a four-sphere saddle subject to small fluctuations. Beyond this, $\mathcal{Z}_{\text{grav}}$ receives contributions from additional geometries that may include Einstein metrics of positive curvature. We discuss how a general positive curvature Einstein metric contributes to $\mathcal{Z}_{\text{grav}}$  at one-loop level. Along the way, we discuss Einstein-Maxwell theory with $\Lambda>0$, and identify an interesting class of closed non-Einstein gravitational instantons. We provide a detailed study for the specific case of $\mathbb{C}P^2$ which is distinguished as the saddle with second largest volume and positive definite tensor eigenspectrum. We present exact one-loop results for scalar particles, Maxwell theory, and Einstein gravity about the Fubini-Study metric on $\mathbb{C}P^2$.

\newpage

\tableofcontents

\section{Introduction}

The quantum entropy of a  de Sitter event horizon, $\mathcal{S}$, was proposed to be macroscopically computed by a gravitational path integral in pioneering work of Gibbons and Hawking \cite{Gibbons:1977mu,Gibbons:1976ue}. Concretely, in four spacetime dimensions,  $\exp{\mathcal{S}}$ is computed by a Euclidean gravitational path integral
\begin{equation}\label{Zgrav11}
\mathcal{Z}_{\text{grav}} = \sum_{\mathcal{M}_4} \int \mathcal{D}g_{\mu\nu} e^{-S_E[g_{\mu\nu}]}~.
\end{equation}
The Einstein-Hilbert action in Euclidean signature reads
\begin{equation}\label{EH}
S_E = \frac{1}{16\pi G} \int d^4 x \sqrt{g}\left(-R +2\Lambda\right)~,
\end{equation}
and is endowed with a positive cosmological constant, $\Lambda>0$. The Newton constant is denoted by $G$. The sum over $\mathcal{M}_4$ in (\ref{Zgrav11}) represents a sum over some subset of closed four-manifolds $\mathcal{M}_4$. The proposal is trivially extended to general spacetime dimensions, and matter fields and higher-derivative corrections can also be straightforwardly  incorporated into $\mathcal{Z}_{\text{grav}}$.
\newline\newline
The path integral, $\mathcal{Z}_{\text{grav}}$,  is difficult to define (in the least) due to the unbounded nature of the conformal mode \cite{Gibbons:1978ac}, as well the absence of a mathematically meaningful sum over closed four-manifolds. Despite these challenges, one can study (\ref{Zgrav11}) in the semiclassical approximation, whereby $G\Lambda\to0^+$. In this limit $\mathcal{Z}_{\text{grav}}$ admits a saddle-point decomposition into a sum over Einstein metrics satisfying $R_{\mu\nu}= \Lambda g_{\mu\nu}$. Moreover, the known set of Einstein four-manifolds is significantly restricted, as reviewed for instance in \cite{anderson2009surveyeinsteinmetrics4manifolds}. Given that any Einstein metric satisfies $R=4\Lambda$, the dominant saddle will be given by the Einstein metric with greatest volume at fixed scalar curvature -- namely, the standard metric on the four-sphere. One then finds, to leading order, that 
\begin{equation}
\log \mathcal{Z}_{\text{grav}} \approx \frac{\Lambda}{8\pi G} V_{S^4} = \frac{3\pi}{G \Lambda}~.
\end{equation}
The above expression is the area of the de Sitter horizon divided by four times the Newton constant $G$, evoking an analogy to the Bekenstein-Hawking relation for black hole entropy. 
\newline\newline
%Gibbons:1979xm,
The one-loop correction about the four-sphere saddle has been studied in \cite{Christensen:1979iy,Volkov:2000ih,Anninos:2020hfj,Law:2020cpj}. It was noted in \cite{hawking2010general,Polchinski:1988ua}, among other places, that due to the unboundedness of the conformal mode, a phase will generally appear in the resulting one-loop contribution. For the $(d+1)$-dimensional sphere, the one-loop phase goes as $(\pm i) \times (\pm i)^{d+2}$, split into a phase $\pm i$ associated to an overall constant mode and a phase $(\pm i)^{d+2}$ associated to the number of non-isometric conformal Killing vector fields \cite{Polchinski:1988ua}. The first part of the phase, coming from the constant part of the conformal mode, has been argued to originate microcanonically  \cite{Anninos:2017hhn,Anninos:2020hfj,Maldacena:2024spf}, while the second part of the phase has been argued to cancel upon incorporating an observer's worldline \cite{Maldacena:2024spf}.\footnote{More generally, the phase is modified by the presence of massless higher spin gauge fields, see equation (5.14) of \cite{Anninos:2020hfj} for a general expression. It was also noted in  \cite{Anninos:2021ihe} that upon coupling a Chern-Simons gauge theory to three-dimensional gravity one can modify and even cancel the overall phase, but this comes at the cost of a parity non-invariant matter sector.} 
\newline\newline
Up to one-loop, the gravitational path integral around the dominant four-sphere saddle in pure gravity admits the following formal expression \cite{Anninos:2020hfj}
\begin{equation}\label{ZdS}
\mathcal{Z}_{\text{grav}} \approx  (\pm i)^{1+5} \frac{e^{\mathcal{S}}}{\text{vol}_{SO(5)}}\left(\frac{8\pi^3}{\mathcal{S}}\right)^{5}  \exp \int_{\mathbb{R}^+}  \frac{dt}{2t} \frac{1+ e^{-t}}{1-e^{-t}} \left( \chi_{\text{bulk}}(t) - \chi_{\text{edge}}(t) \right)~,
\end{equation}
where the bulk and edge characters are given by
\begin{equation}
\chi_{\text{bulk}}(t) = \frac{10 e^{-3t} - 6 e^{-4t}}{\left(1-e^{-t}\right)^3}~, \quad\quad \text{and} \quad\quad \chi_{\text{edge}}(t) = \frac{10 e^{-2t} - 2 e^{-3t}}{\left(1-e^{-t}\right)}~.
\end{equation}
The bulk character, $\chi_{\text{bulk}}$, is an $SO(1,4)$  group character of Harish-Chandra's discrete series unitary irreduble representation carried by the single-particle Hilbert space of the quantised graviton \cite{Higuchi:1991tn} in de Sitter space (see also \cite{hirai,Basile:2016aen,Sun:2021thf} for a general overview of the de Sitter representation theory). The edge character, $\chi_{\text{edge}}$, has an associated ultraviolet divergence that is codimension two, and has been explored in terms of a putative edge mode physics in \cite{David:2022jfd,Law:2025ktz}. The one-loop contribution can be regularised using standard heat kernel methods, and the finite part of the regularised expression can be succinctly expressed in terms of generalised zeta functions. 
\newline\newline
Of the known Einstein metrics with positive curvature, the one with the second largest volume is the Fubini-Study metric on $\mathbb{C}P^2$. This spacetime was interpreted as a gravitational instanton in \cite{Eguchi:1976db,Gibbons:1978zy}.  It is noted in that work, as well as in \cite{Hawking:1978ghb}, that $\mathbb{C}P^2$ does not admit a spin structure. This is due to a non-trivial second Stiefel-Whitney class. Nonetheless, like all orientable four-manifolds \cite{hirzebruch1958felder,teichner1994all}, $\mathbb{C}P^2$ admits a generalised spin-c structure \cite{Hawking:1977ab,Pope:1980ub}.\footnote{As such, provided the spectrum of fermions carries judiciously chosen $U(1)$ gauge charges, they can be placed on $\mathbb{C}P^2$. For instance, the standard model equipped with sterile right handed neutrinos and a gauged $U(1)_{B-L}$ symmetry can be placed on a manifold with a spin-c structure \cite{Garcia-Etxebarria:2018ajm,Davighi:2022icj}. This remains true for a Higgsed $U(1)_{B-L}$, whereby the low energy effective field theory will also contain Abelian-Higgs vortices \cite{Abrikosov:1956sx,Nielsen:1973cs}. For a discussion of mixed gravitational anomalies associated to the $U(1)_{B-L}$ symmetry, see \cite{Putrov:2023jqi}. For the mixed gravitational anomaly to be present, one must have that the number of sterile right handed neutrinos differs from the number of generations in the standard model. Further discussion on $U(1)_{B-L}$ can be found in \cite{Heeck:2014zfa}. Potential phenomenological links between $\Lambda>0$ and $U(1)_{B-L}$ are worth further exploration.} In addition, there are several other Einstein metrics including $S^2 \times S^2$, which has been interpreted in terms of the Euclidean continuation of a Nariai black hole \cite{Ginsparg:1982rs}. It is worth contrasting the multiple saddle-point contributions to $\mathcal{Z}_{\text{grav}}$ with the case of asymptotically flat and anti-de Sitter space in Euclidean signature. In the flat case, the positive action theorem \cite{Schon:1979uj,Gibbons:1979xn} indicates that the standard flat metric is the unique asymptotically flat solution to the Euclidean Einstein equations on an $\mathbb{R}^4$ manifold. In Euclidean AdS$_4$, upon fixing the asymptotic metric to be conformally equivalent to the flat metric on $\mathbb{R}^3$, one generically expects the unique solution to the Einstein equations with $\Lambda<0$ to be Euclidean AdS$_4$, given that the dual CFT$_3$ has a unique conformally invariant vacuum. 
\newline\newline
From a technical viewpoint, the purpose of this article is to calculate some properties around the $\mathbb{C}P^2$ saddle of Einstein gravity with $\Lambda>0$. We consider the case of scalar fields, the Maxwell field, and the linearised graviton. We show that many of the one-loop techniques developed for the $(d+1)$-dimensional sphere in \cite{Anninos:2020hfj} can be readily ported to $\mathbb{C}P^2$. From a physical point of view, of the known Einstein metrics with positive scalar curvature, the  $\mathbb{C}P^2$ saddle constitutes the leading subdominant saddle to $\mathcal{Z}_{\text{grav}}$. Aside from the four-sphere saddle, $\mathbb{C}P^2$ is the only one that is known to have a strictly positive tensor mode eigenspectrum. The $\mathbb{C}P^2$ saddle, like any non-spherical Euclidean saddle, breaks the de Sitter isometries. It would be interesting to understand the implications of this for  physical observables in $\Lambda>0$ quantum gravity, perhaps along the lines of \cite{Goheer:2002vf}. We also note that the one-loop contribution of the $S^2\times S^2$ Euclidean saddle was computed in \cite{Volkov:2000ih}, which exhibits a negative mode in the tensor eigenspectrum. 
\newline\newline
The outline of the paper is as follows. In section \ref{geometry}, we discuss some general properties of Einstein metrics, with an emphasis on $\mathbb{C}P^2$. We consider the general structure of $\mathcal{Z}_{\text{grav}}$ in (\ref{Zgrav11}) at one-loop, and comment on the case of Einstein-Maxwell theory with $\Lambda>0$. In section \ref{spectra}, we review the spectra of various Laplace type operators on $\mathbb{C}P^2$ as computed in  \cite{Warner:1982fs,Pope:1980ub,Vassilevich:1993yt,boucetta2010spectra}. In section \ref{scalar}, we compute the one-loop partition function of a scalar field on $\mathbb{C}P^2$. We express the result in a form similar the character formulae in \cite{Anninos:2020hfj}. In section \ref{maxwell}, we compute the one-loop partition function for the Maxwell field on $\mathbb{C}P^2$. As a check on the spectra, we compare the logarithmic divergence to that computed by a standard heat kernel coefficient analysis \cite{Vassilevich:2003xt}. In section \ref{Zgrav1sec}, we compute the one-loop partition function for linearised Einstein gravity around a $\mathbb{C}P^2$ saddle.  As a check on the spectra, we compare the resulting logarithmic divergence to that computed by a standard heat kernel coefficient analysis (see for example \cite{Christensen:1979iy}, and \cite{Vassilevich:1993yt} for the spefic case of $\mathbb{C}P^2$). A successful comparison requires a careful treatment of the zero mode sector that arises upon dividing by the locally defined volume of the diffeomorphism group. In appendix \ref{scalarZapp}, we test our scalar partition function directly from a Green's function analysis. In appendix \ref{integrals}, we provide explicit expressions for the heat kernel integrals that appear throughout the main text. In appendix \ref{S2S2app}, we provide details regarding the one-loop gravitational partition function about an $S^2\times S^2$ saddle. 
%\newline\newline

\section{Geometric features}\label{geometry}

In this section we will discuss properties of smooth solutions to the four-dimensional Einstein equations with positive cosmological constant in Euclidean signature
\begin{equation}\label{Einstein}
R_{\mu\nu} -\frac{1}{2} g_{\mu\nu} R + \Lambda g_{\mu\nu} = 0~.
\end{equation}
The solutions are known as Einstein metrics and have constant positive Ricci scalar, $R=4\Lambda$, along with $R_{\mu\nu} = \Lambda g_{\mu\nu}$ by virtue of (\ref{Einstein}).  The Einstein metrics reside on a given closed four-manifold, $\mathcal{M}_4$. The only known smooth Einstein metric on a topological four-sphere is the round four-sphere, whose volume is given by $V_{S^4} = \tfrac{24\pi^2}{\Lambda^2}$. The other known four-manifolds admitting smooth Einstein metrics are $\mathbb{C}P^2$, $S^2\times S^2$, and $\mathbb{C}P^2 \# k\overline{\mathbb{C}P^2}$, with $1\le k \le 8$.\footnote{The connected sums $\mathbb{C}P^2 \# k\overline{\mathbb{C}P^2}$ are obtained by removing $k$ balls from $\mathbb{C}P^2$ and gluing it to another excised $\mathbb{C}P^2$ of opposite orientation at $k$ generic points. For $k=1$ there a is known smooth Einstein metric with constant curvature identified by Page in \cite{Page:1978vqj}. It describes an $S^2$ non-trivially fibered over another $S^2$, and admits four Killing vector fields.  This metric has smaller volume than the $S^2\times S^2$ saddle. For $k=2$ an Einstein metric was identified by Chen, Le Brun, and Weber \cite{chen2008conformally}. For $k\ge5$ the space of Einstein metrics on $\mathbb{C}P^2 \# k\overline{\mathbb{C}P^2}$ admits a non-trivial moduli space.} A survey can be found in \cite{anderson2009surveyeinsteinmetrics4manifolds}.  Of these, only $S^4$ and $S^2\times S^2$ admit a standard spin-structure. However, the remaining four-manifolds admit a spin-c structure \cite{Hawking:1977ab}. As shown in \cite{Garcia-Etxebarria:2018ajm}, the spin-c structure implies that the standard model can be placed on this manifold, provided that the $B-L$ symmetry, where $B$ is the baryon number and $L$ is the lepton number, is gauged. In what follows, we assume that any matter theory coupled to our gravitational action is compatible with a spin-c structure. 
% (page 25)
\newline\newline
Of the known Einstein metrics with a given positive curvature, the round four-sphere has the largest volume, followed by the Fubini-Study metric on $\mathbb{C}P^2$. It is topologically distinct from the four-sphere as can be  seen from the Euler characteristic $\chi_{S^4}=2$, $\chi_{\mathbb{C}P^2}=3$, $\chi_{S^2\times S^2}=4$, and $\chi_{\mathbb{C}P^2 \# k\overline{\mathbb{C}P^2}}=3+k$,\footnote{To see this, recall that the Euler characteristic of a connected sum of two four-manifolds $\mathcal{M}_4$ and $\mathcal{N}_4$ satisfies $\chi_{\mathcal{M}_4 \# \mathcal{N}_4} = \chi_{\mathcal{M}_4}+\chi_{\mathcal{N}_4}- \chi_{S^4}$.} with $1\le k \le 8$ (see for example \cite{anderson2009surveyeinsteinmetrics4manifolds}). The $\mathbb{C}P^2 \# k\overline{\mathbb{C}P^2}$ further have $b_2 = 1+k$ independent two-cycles. The Hirzebruch signature $\tau_{\mathcal{M}_4}$ for a given orientation is given by $\tau_{S^4}=0$, $\tau_{\mathbb{C}P^2}= -1$, $\tau_{S^2\times S^2}=0$, and $\tau_{\mathbb{C}P^2 \# k\overline{\mathbb{C}P^2}}=k-1$, as listed in \cite{Gibbons:1979xm}. We note that these all satisfy the inequality $\chi_{\mathcal{M}_4} > \tfrac{3}{2} |\tau_{\mathcal{M}_4}|$ obeyed by all non-Ricci flat Einstein metrics \cite{hitchin1974compact,thorpe1969some}. For $k\ge 9$, the inequality would be violated. We further note that all these manifolds have vanishing first Betti number. Indeed, it follows from Myer's theorem \cite{myers1941riemannian}, that any space with bounded Ricci curvature can only have a finite  fundamental group. 
\newline\newline
The Einstein-Hilbert action (\ref{EH}) evaluated on an Einstein metric is given by
\begin{equation}\label{SV}
-\mathcal{S}_{\mathcal{M}_4} \equiv \frac{1}{16\pi G} \int d^4 x\sqrt{g} \left(-R +2 \Lambda\right) = -\frac{\Lambda}{8\pi G} V_{\mathcal{M}_4}~.
\end{equation}
There is no boundary term for the action because we are taking the manifold to be closed. Of the known Einstein metrics, following the four-sphere, $\mathbb{C}P^2$ has the largest on-shell Euclidean action. 
\newline\newline
Finally, although we only consider four-dimensions here, it is worth pointing out that in other dimensions the number of known Einstein manifolds is infinite. For instance, in three spacetime dimensions, one has the infinite family of Lens spaces that are given by smooth quotients of $S^3$. The sum over such three-dimensional topologies was examined in \cite{Carlip:1992wg,Guadagnini:1994ahx,Castro:2011xb}. In five-dimensions, and higher, product metrics of the type $S^3\times S^p$ are generally present. As such, we automatically have an infinite class of Einstein manifolds. Moreover, in $(2n+3)$-spacetime dimensions, with $n \in \mathbb{Z}^+$, one finds increasigly large families of smooth Einstein metrics on $S^{2n+1}$ \cite{boyer2005einstein}. For instance, there is an infinite family of B\"ohm metrics on $S^5$ \cite{bohm1998inhomogeneous} enjoying an $SO(3)\times SO(3)$ isometry.

\subsection{Fubini-Study metric on $\mathbb{C}P^2$}

The Fubini-Study metric on $\mathbb{C}P^2$ can be written in terms of a pair of complex coordinates $\zeta^i = (\zeta^1,\zeta^2) \in \mathbb{C}^2$. As $\mathbb{C}P^2$ is a K\"ahler manifold, we can derive the metric from a K\"ahler potential $K$, 
\begin{equation}\label{FS}
K = \frac{6}{\Lambda} \log \left( 1+ \frac{\Lambda}{6} \delta_{i \bar{j}} \zeta^i \bar{\zeta}^{\bar{j}} \right)~, \quad\quad g_{i\bar{j}} = \partial_{\zeta^i} \partial_{\bar{\zeta}^{\bar{j}}} K~,
%, \quad\quad z^i \in \mathbb{C}~,
%(z^i,\bar{z}^{\bar{i}}) 
\end{equation}
with $i \in (1,2)$. The Fubini-Study metric admits eight isometries. This follows from the fact that it is a coset space $\tfrac{U(3)}{U(1)\times U(2)}$. Concretely, the isometry group of the Fubini-Study metric on $\mathbb{C}P^2$ is $\tfrac{SU(3)}{\mathbb{Z}_3}$. Consequently, the Fubini-Study metric is not maximally symmetric. Moreover, by a theorem of Yano and Nagano \cite{yano1959einstein}, the Fubini-Study metric admits no non-isometric conformal Killing vector fields. Another often employed coordinate system for the Fubini-Study metric is as follows
\begin{equation}\label{taubNUT}
ds^2 = \frac{3}{2\Lambda} \left( \frac{4 dr^2}{(1+r^2)^2}  + \frac{r^2}{1+r^2}\left(\omega_1^2+\omega^2_2\right) + \frac{r^2}{(1+r^2)^2} \omega_3^2  \right)~,
\end{equation}
with $r\in\mathbb{R}^+$ and the one-forms $\omega_1$, $\omega_2$, $\omega_3$ being those of the standard three-sphere. Explicitly, the three one-forms are given by
\begin{eqnarray}
\omega_1 &\equiv& \cos\psi d\theta + \sin\psi\sin\theta d\phi~, \\
\omega_2 &\equiv& \sin\psi d\theta - \cos\psi\sin\theta d\phi~, \\
\omega_3 &\equiv&  d\psi+\cos\theta d\phi~,
\end{eqnarray}
with $\psi \in (0,4\pi)$, $\theta \in (0,\pi)$, and $\phi \sim \phi+2\pi$. The metric at a constant $r$ slice is given by a squashed three-sphere. We can view the metric (\ref{taubNUT}) as a Euclidean version of the Taub-NUT de Sitter cosmology at complexified parameters.
\newline\newline
The volume of the Fubini-Study metric is given by $V_{\mathbb{C}P^2} = \tfrac{18\pi^2}{\Lambda^2}$ which is less than $V_{S^4} = \tfrac{24\pi^2}{ \Lambda^2}$, but greater than $V_{S^2\times S^2} = \tfrac{16\pi^2}{\Lambda^2}$. It is also known \cite{Page:1978vqj} that $V_{\mathbb{C}P^2 \# \overline{\mathbb{C}P^2}} \approx \tfrac{150.9}{ \Lambda^2}$, that $V_{\mathbb{C}P^2 \# 2 \overline{\mathbb{C}P^2}} \approx \tfrac{132.8}{ \Lambda^2}$ \cite{hall2014spectrum}, and that \cite{doran2008numerical}
%< V_{S^2\times S^2}
\begin{equation}
V_{\mathbb{C}P^2 \# k \overline{\mathbb{C}P^2}} = (9-k)\frac{2\pi^2}{\Lambda^2}~, \quad\quad k =3,\ldots,8~.
\end{equation}
The above equation follows from the fact that the K\"ahler form of a K\"ahler-Einstein metric is simply related to the Ricci form and consequently the first Chern class. Based on the general expression (\ref{SV}), the on-shell action of the Fubini-Study metric on $\mathbb{C}P^2$ is given by
\begin{equation}
\mathcal{S}_{\mathbb{C}P^2} = \frac{9\pi}{4 G\Lambda}~,
\end{equation}
and one can similarly proceed with the our Einsten spaces. We note that
\begin{equation}\label{orderS}
\mathcal{S}_{{S^4}}>\mathcal{S}_{\mathbb{C}P^2}>\mathcal{S}_{{S^2\times S^2}} > \mathcal{S}_{{\mathbb{C}P^2 \# \overline{\mathbb{C}P^2}}}  > \mathcal{S}_{{\mathbb{C}P^2 \# 2\overline{\mathbb{C}P^2}}} > ... > \mathcal{S}_{{\mathbb{C}P^2 \# 8 \overline{\mathbb{C}P^2}}}~.
\end{equation}
Thus, if $\mathbb{C}P^2$ contributes to physical processes, it dominates over $S^2\times S^2$, and most likely over all other Einstein four-manifolds except $S^4$. 

%We are not aware of results on $V_{\mathbb{C}P^2 \# 2 \overline{\mathbb{C}P^2}}$, but perhaps based on the pattern in (\ref{orderS}) it will reside somewhere in the interval $[118.4,150.9]$, in units where $\Lambda=1$.
%\newline\newline
% and $\text{dim} \, U(3)=8$

\subsection{Pieces of the partition function of a $\Lambda>0$ Universe}

Based on the known Einstein metrics, and assuming they contribute to the gravitational path integral, we have the following approximate expression for $\mathcal{Z}_{\text{grav}}$, the partition function of the universe
\begin{multline}\label{ZU}
\mathcal{Z}_{\text{grav}} \approx i \left(i^5 \mathcal{S}_{{S^4}}^{-5} \exp {\mathcal{S}_{{S^4}}}  +  \mathcal{S}_{\mathbb{C}P^2}^{-4} \exp{\mathcal{S}_{\mathbb{C}P^2}} - i  \mathcal{S}_{S^2\times S^2}^{-3} \exp{\mathcal{S}_{S^2\times S^2}}   \right. \\  \left.  - i  \mathcal{S}_{{\mathbb{C}P^2 \# \overline{\mathbb{C}P^2}}}^{-2}  \exp \mathcal{S}_{{\mathbb{C}P^2 \# \overline{\mathbb{C}P^2}}}  + \ldots \right)~.
\end{multline}
We have included both the leading contribution from the on-shell action $\mathcal{S}_{\mathcal{M}_4}$ for the given Einstein metric, and a piece that comes from dividing by the appropriately normalised volume of the isometry group, $G_4$, of $\mathcal{M}_4$ \cite{Volkov:2000ih,Anninos:2020hfj}. Our choice of phase complies with the discussion of \cite{Ivo:2025yek}. Generally, we should sum over both canonical orientiations of the given four-manifold when it does not admit orientation reversing diffeomorphisms. Provided the Einstein metric is not accompanied by a moduli space of solutions, this contribution will generally go as
\begin{equation}
\log \mathcal{Z}^{(1)}_{\text{grav}} \propto -\frac{\dim G_4}{2} \log \mathcal{S}_{\mathcal{M}_4}~.
\end{equation}
We will verify this explicitly for $\mathbb{C}P^2$ below. The overall $i$ in (\ref{ZU}) comes from the constant part of the conformal mode in the metric fluctuation. The $i^5$ from the $S^4$ contribution comes from the five non-isometric conformal Killing vector fields \cite{Polchinski:1988ua}. The additional $-i$ in ${S^2\times S^2}$ and ${\mathbb{C}P^2 \# \overline{\mathbb{C}P^2}}$ stems from negative modes in the tensor eigenspectrum \cite{Volkov:2000ih,Young:1983wwy,Hennigar:2024gbg,Shi:2025amq,Ivo:2025yek}. It is not known whether the Einstein metric of Chen-LeBrun-Weber \cite{chen2008conformally} has negative tensor modes, an analysis of various spectra can be found in \cite{hall2014spectrum}.\footnote{It would be interesting to clarify the appearance of such phases beyond one-loop (or their potential one-loop exactness), as analysed, for instance, in \cite{Anninos:2021ene}.} However, it is known that the Chen-LeBrun-Weber \cite{chen2008conformally} has only two isometries, and so would contribute the following term to $\mathcal{Z}_{\text{grav}}$
\begin{equation}
a_0 \,  \mathcal{S}_{{\mathbb{C}P^2 \# 2 \overline{\mathbb{C}P^2}}}^{-1}  \exp \mathcal{S}_{{\mathbb{C}P^2 \# 2 \overline{\mathbb{C}P^2}}}~,
\end{equation}
where $a_0\in \mathbb{C}$ is either pure imaginary or real, and $\mathcal{S}_{{\mathbb{C}P^2 \# 2 \overline{\mathbb{C}P^2}}}=\frac{\Lambda}{8\pi G} V_{\mathbb{C}P^2 \# 2\overline{\mathbb{C}P^2}}$ is finite and real. There will also be contributions to $\mathcal{Z}_{\text{grav}}$ from the matter fields about each saddle.
\newline\newline
Interestingly, for $5\le k\le 8$, the manifolds $\mathbb{C}P^2 \#  k \overline{\mathbb{C}P^2}$ admit a K\"ahler-Einstein metric and have a moduli space of non-vanishing dimension \cite{tian1987kahler}, whilst admitting no isometries. As such, their volume is determined topologically in terms of their first Chern class, as discussed in \cite{tian1990calabi}. It is also shown in \cite{tian1987kahler} that the complex dimension of this moduli space satisfies $\dim_{\mathbb{C}}\mathfrak{M}_k \ge (k-4)$. The presence of a moduli space suggests that the contributions from these four-manifolds to $\mathcal{Z}_{\text{grav}}$ go as
\begin{equation}
a_k \, \left( \mathcal{S}_{{\mathbb{C}P^2 \# k \overline{\mathbb{C}P^2}}} \right)^{\dim_{\mathbb{R}}\mathfrak{M}_k}  \exp \mathcal{S}_{{\mathbb{C}P^2 \# k \overline{\mathbb{C}P^2}}}~, \quad\quad k =5,\ldots,8~,
\end{equation}
where the $a_k\in \mathbb{C}$ are either pure imaginary or real, and will go as the volume of $\mathfrak{M}_k$. Notice that now the power-law contribution has a positive power. 
\newline\newline
The general lesson is that $\mathcal{Z}_{\text{grav}}$, and its underlying ultraviolet completion, may exhibit a rich mathematical structure that could significantly restrict candidate microphysical theories. Around each of the leading saddle-point contributions comes a quantum contribution from the matter fields as well as the one-loop contribution of the gravitational fluctuations themselves. The gravitational one-loop contributions for $S^4$ and $S^2\times S^2$ have been studied in detail \cite{Volkov:2000ih,Anninos:2020hfj}. In later sections we will consider the case of $\mathbb{C}P^2$. 
\newline\newline
Finally, we should note that there can be additional gravitational couplings in the theory due to higher derivative terms. For instance, there could be a coupling $\vartheta$ associated to the Euler characteristic, $\chi_{\mathcal{M}_4}$, which can be expressed locally in terms of the Gauss-Bonnet invariant. For an Einstein metric satisfying $R_{\mu\nu} = \Lambda g_{\mu\nu}$, a closed manifold satisfies
\begin{equation}
\chi_{\mathcal{M}_4} = \frac{1}{32\pi^2} \int d^4 x\sqrt{g} R_{\mu\nu\rho\sigma} R^{\mu\nu\rho\sigma}~,
\end{equation}
and we have already listed some values of $\chi_{\mathcal{M}_4}$. We list them again here for convenience, $\chi_{S^4}=2$, $\chi_{\mathbb{C}P^2}=3$, $\chi_{S^2\times S^2}=4$, and $\chi_{\mathbb{C}P^2 \# k\overline{\mathbb{C}P^2}}=3+k$, with $1\le k \le 8$ . The coupling will add further structure of $\mathcal{Z}_{\text{grav}}$, essentially multiplying each term by a factor $e^{-\vartheta \chi_{\mathcal{M}_4} }$. Evidently, to suppress topological contributions (with positive Euler characteristic) one can take $\vartheta \gg 1$. Alternatively, non-spherical topologies can be enhanced upon taking $\vartheta\ll -1$.  Similar considerations apply for the Hirzebruch signature,
\begin{equation}
\tau_{\mathcal{M}_4} = \frac{1}{96\pi^2} \int d^4 x \sqrt{g} \varepsilon^{\rho\sigma\alpha\beta} R_{\mu\nu\rho\sigma} {R^{\mu\nu}}_{\alpha\beta}~.
\end{equation}
%, which could  for instance appear from fermionic loops \cite{eguchi}. 
The sign of  $\tau_{\mathcal{M}_4}$ depends on the choice of orientation. Much like the $\theta$ term of electromagnetism, the contribution from $\tau_{\mathcal{M}_4}$ will appear as a phase of the Euclidean path integral (see \cite{Ryu:2010ah} for the appearance of such a phase in topological insulators and superconductors). 
%In general, we will need to include $\mathcal{M}_4$ with both orientations to the partition function.  

\subsection{Einstein-Maxwell-$\Lambda$ Theory} 

As a slightly more general theory, we can consider Einstein Maxwell theory with $\Lambda>0$ with action
\begin{equation}
S_{EM} = \frac{1}{16\pi G}\int d^4 x \sqrt{g} \left( -R +2 \Lambda + F_{\mu\nu} F^{\mu\nu} \right)~,
\end{equation}
where $F = d A$ is the electromagnetic field strength associated to an Abelian $U(1)$ gauge field $A_\mu$. This theory can now have additional instantons \cite{Brill:1992ce}. A simple example is given by the product space $H^2 \times S^2$. This is the Euclidean continuation of the near horizon of a electromagnetically charged extremal black hole in de Sitter space. For the case of a purely magnetic charge, the metric and gauge field strength are given by  \cite{romans1992supersymmetric,mann1995cosmological,Booth:1998gf,Hawking:1995ap}
\begin{equation}\label{EMsol}
ds^2 = \frac{1}{A_1} dH_2 + \frac{1}{A_2} \left(d\theta^2 + \sin^2\theta d\phi^2\right)~, \quad\quad F = Q \sin \theta d\theta \wedge d\phi~,
\end{equation}
where $H_2$ is the standard hyperbolic metric with Ricci scalar $R_{H^2} = -2$. For the above to solve the Einstein-Maxwell equations, we must further take
\begin{equation}
A_2 = \frac{\sqrt{1-4 \Lambda  Q^2}+1}{2 Q^2}~, \quad\quad A_1 = A_2 -  2\Lambda~.
\end{equation}
We have selected the root that yields a non-vanishing answer for $A_2$ as $\Lambda\to 0$.  We also require that $Q^2 \in (0, \tfrac{1}{4\Lambda})$, such that $A_1>0$. The flux quantisation condition is such that $Q = \tfrac{\mathfrak{m}}{e} \sqrt{\pi G}$, with $\mathfrak{m}\in\mathbb{Z}$, and $e$ the $U(1)$ gauge coupling. A similar metric can be obtained for the black hole with both electric and magnetic fluxes, but we relegate a complete study to future work.
\newline\newline
We can quotient the hyperbolic space, $H_2$, by a freely acting subgroup of $SL(2,\mathbb{R})$ to yield a smooth closed hyperbolic surface, $\Sigma_h$, of genus $h$. The resulting Euclidean instanton is then closed, with finite four-volume $V^{(h)}_Q$. Minus the Euclidean action is given by
\begin{equation}
\mathcal{S}^{(h)}_Q = \frac{V^{(h)}_Q}{8\pi G} \left( \Lambda -  Q^2 A_2^2 \right)~.
\end{equation}
For a genus $h$ surface, the four-volume of (\ref{EMsol}) can be computed, using the Gauss-Bonnet theorem, and is given by
\begin{equation}
V^{(h)}_Q = \frac{16\pi^2}{A_1 A_2} \times (h-1)~, \quad\quad h = 2,3,4,\ldots
\end{equation}
Thus, we are left with
\begin{equation}
\mathcal{S}^{(h)}_Q = \frac{4 \pi  Q^2 (1-h)}{G\left( 1+ \sqrt{1-4 \Lambda  Q^2} \right)}~.
\end{equation}
%We note that setting $h=0$ in the above formula yields the correct on-shell action for the magnetically charged $S^2 \times S^2$ Nariai saddle, as computed in section III of \cite{Hawking:1995ap}.
%\newline\newline
It would be interesting to understand the contribution to the path integral from the sum of all genera. Each Riemann surface will come with a moduli space whose volume is given by the Weil–Petersson metric.\footnote{This is reminiscent of computations in two-dimensional quantum gravity, where the resulting sum is asymptotic leading to the inclusion of manifolds with boundaries \cite{Polchinski:1994fq,Ginsparg:1991ws,Eynard:1992sg,Shenker:1990uf}. Perhaps the same phenomenon occurs for the Gibbons-Hawking path integral, as envisioned in \cite{Anninos:2021eit,Anninos:2024wpy,Silverstein:2024xnr,banihashemi2025flat}. Other aspects of higher topology in de Sitter space, from a wavefunctional perspective include \cite{Anninos:2012ft,Banerjee:2013mca,Cotler:2019nbi,Maldacena:2019cbz,Anninos:2022ujl,Aguilar-Gutierrez:2023ril,Collier:2025lux,Turiaci:2025xwi,Blacker:2025zca}.}
There is also a solution given by $\mathbb{R}^2 \times S^2$, namely
\begin{equation}
ds^2 =  \left(dx^2 + dy^2\right) + \frac{1}{A_2} \left(d\theta^2 + \sin^2\theta d\phi^2\right)~, \quad\quad F = Q \sin \theta d\theta \wedge d\phi~, 
\end{equation}
with $Q^2 = \tfrac{1}{4\Lambda}$, and $A_2 = \tfrac{1}{2Q^2}$. We can quotient the $\mathbb{R}^2$ to a torus of complex structure $\tau$. The metric can be viewed as the Euclidean counterpart of the near horizon limit of the ultracold black hole in de Sitter space. The on-shell action now vanishes for all tori! Nonetheless, there is a moduli space of tori that must be integrated over in the path integral, that may lead to mild divergences (reminiscent again of two-dimensional quantum gravity on a torus \cite{Bershadsky:1990xb}). This issue, for the more general case with both electric and magnetic charge, along with the higher genus contributions, will be analysed elsewhere. 
%\begin{equation}
%-\mathcal{S}^{(h)}_Q = \frac{V^{(0)}_Q}{8\pi G} \left( \Lambda -  \frac{1}{4 Q^2} \right)~.
%\end{equation}
\newline\newline
There will also be Euclidean versions of the charged Nariai black hole, including the case with rotation \cite{Hawking:1995ap}. Moreover, if $\mathcal{M}_4$ admits a K\"ahler metric with constant Ricci scalar, one always has a solution to the $\Lambda>0$ Einstein-Maxwell equations \cite{Flaherty:1978wh,lebrun2016einstein}, and the Maxwell field strength is built entirely from the geometric K\"ahler structure. Finally,  \cite{lebrun2016einstein} also uncovered an infinite family of Einstein-Maxwell-$\Lambda$ solutions on $S^2\times S^2$ and $\mathbb{C}P^2 \# \overline{\mathbb{C}P^2}$ with a metric which is only conformally K\"ahler. The effective  Einstein-Maxwell-$\Lambda$ field theory appears with infinitely many new couplings, as compared to the Einstein theory. It is perhaps reassuring, then, that there are also an infinite number of additional saddles to accompany these.
% (page 302). 

%$\mathcal{S}_{{S^4}}>\mathcal{S}_{\mathbb{C}P^2}>\mathcal{S}_{{S^2\times S^2}} > \mathcal{S}_{{\mathbb{C}P^2 \# \overline{\mathbb{C}P^2}}} $

\section{Laplace type spectra of $\mathbb{C}P^2$}\label{spectra}

In this section we review the eigenspectra of a variety of Laplace type operators on $\mathbb{C}P^2$.  The most efficient method to compute spectra of quadratic differential operators on the Fubini-Study metric on $\mathbb{C}P^2$ uses the fact that it is related to the round five-sphere via a Hopf fibration. In particular, $S^5$ can be viewed as an $S^1$ non-trivially fibered over a $\mathbb{C}P^2$ base space. As such, one can deduce the spectra of Laplacian operators directly from those on $S^5$ which are well known \cite{rubin1984eigenvalues}. The task entails decomposing a subspace of irreducible representations of the $SU(4)$ (double cover) isometry group of $S^5$ into irreducible representations of the $SU(3)$ isometry group of $\mathbb{C}P^2$ \cite{macfarlane1975generalization}.  This subspace is chosen so as to be independent of the Killing isometry associated to the fibre direction. This was undertaken in \cite{Warner:1982fs,Pope:1980ub,Vassilevich:1993yt,Vassilevich:1995bg,boucetta2010spectra}. For a spin-$s$ field, the underlying $SU(4)$ irreps relevant to the $SU(3)$ decomposition have weights $(p,2n,p)$ with $p=0,1,\ldots,s$, with dimension
\begin{equation}
\dim_{SU(4)}({p, 2n,p}) = \frac{1}{12} (2 n+1) (p+1)^2 (2 n+p+2)^2 (2 n+2 p+3)~.
%\frac{1}{12}(1+n_1)(1+n_2)(2+n_1+n_2)~.
\end{equation}
We further recall that the dimension  of the irrep of $SU(3)$ labelled by the weights $(n_1,n_2)$ is given by \cite{coleman1965fun}
\begin{equation}
\dim_{SU(3)}({n_1, n_2}) =\frac{1}{2}(1+n_1)(1+n_2)(2+n_1+n_2)~.
\end{equation}

%We list the results below. 

\subsection{Scalar Laplacian spectrum} 

For the scalar Laplacian, $\Delta_{(0)} = -\nabla^2$, the spectra and degeneracies are given by\footnote{We have corrected a minor typographical error in equation (6.2) of \cite{Warner:1982fs}.}
\begin{equation}\label{scalarspec}
\lambda^{(0)}_n = \frac{2\Lambda}{3}n(n+2)~, \quad\quad d^{(0)}_n = (n+1)^3~,  \quad\quad n \in \mathbb{N}~.
\end{equation}
The eigenmode $\lambda^{(0)}_0$ with vanishing eigenvalue corresponds to the constant mode. The eigenmodes are associated with the $(n,n)$ irrep of $SU(3)$.

\subsection{Hodge-De Rham operator spectrum} 

For the Hodge-De Rham operator acting on transverse vector fields, $\Delta_{(1)}$, the spectra and degeneracies are given by
\begin{eqnarray}\label{vecspec}
\lambda^{(1)}_n &=& \frac{2\Lambda}{3}n(n+2)~, \quad\quad d^{(11)}_n = (n+1)^3~,  \quad\quad n \in \mathbb{Z}^+~, \\ \label{vecspec2}
\gamma^{(1)}_n &=& \frac{2\Lambda}{3}(n+1)(n+2)~, \quad\quad d^{(12)}_n = n(n+3)(2n+3)~, \quad\quad n \in \mathbb{Z}^+~.
\end{eqnarray}
The eight $\lambda^{(1)}_1$ eigenmodes are the Killing vector fields of the Fubini-Study metric. We note that $\Delta_{(1)}=-\nabla^2_{(1)} +\Lambda$, where $\nabla^2_{(1)}$ is the Laplacian acting on transverse vector fields. The operator $\Delta_{(1)}$  acts on the transverse part of the gauge field in Maxwell theory on an Einstein metric background satisfying $R_{\mu\nu} = \Lambda g_{\mu\nu}$. The eigenmodes are associated with the $(n+1,n+1)\oplus (n-1,n+2)\oplus (n+2,n-1)$ irreps of $SU(3)$. 

\subsection{Lichnerowicz operator spectrum} 

For the transverse-traceless part of the Lichnerowicz operator, $\Delta_{(2)}$, the spectra and degeneracies are given by
\begin{eqnarray}\label{lichspec}
\lambda^{(2)}_n &=& \frac{2\Lambda}{3}(n+2)(n+4)~, \quad\quad d^{(21)}_n = (n+3)^3~,  \quad\quad n \in \mathbb{N}~, \\ \label{lichspec2}
\gamma^{(2)}_n &=& \frac{2\Lambda}{3}(n+3)(n+4)~, \quad\quad d^{(22)}_n = (2n+7)(n+2)(n+5)~, \quad\quad n \in \mathbb{N}~, \\ \label{lichspec3}
\delta^{(2)}_n &=& \frac{2\Lambda}{3}(n^2+8n+18)~, \quad\quad d^{(23)}_n = 2(n+1)(n+4)(n+7)~, \quad\quad n \in \mathbb{N}~. 
\end{eqnarray}
We note that $\Delta_{(2)} h_{\mu\nu} = - \nabla^2_{(2)} h_{\mu\nu} -2R_{\mu\nu\alpha\beta} h^{\alpha\beta}+ 2\Lambda h_{\mu\nu}$, where $\nabla^2_{(2)}$ is the Laplacian acting on transverse-traceless symmetric rank two tensors. The shifted operator $\Delta_{(2)}-2\Lambda$ appears in the linearised Einstein theory with $\Lambda>0$, expanded around an Einstein metric satisfying $R_{\mu\nu}=\Lambda g_{\mu\nu}$. The eigenmodes are associated with the $(n+2,n+2)\oplus (n+1,n+4)\oplus (n+4,n+1)\oplus (n,n+6)\oplus (n+6,n)$ irreps of $SU(3)$.\footnote{We have corrected a minor typographical error in the left most column of Table 1 of \cite{Warner:1982fs}.}
\newline\newline
An alternative way of obtaining the spectra is in terms of a Kaluza-Klein reduction problem with a $U(1)$ gauge field \cite{Pope:1980ub}, and finding appropriately neutral solutions to the wave-equations. The $U(1)$ gauge field in question is the K\"ahler one-form of $\mathbb{C}P^2$. 
\newline\newline
As a final note, we remark that we can further decompose the $SU(3)$ irreducible representations into $SU(2)\times U(1)$ irreducible representations. Following \cite{hagen1964triality}, we have that the $SU(3)$ irrep with highest weight $(n_1,n_2)$ breaks into the $SU(2)$  irreducible representations of dimension $(m_1-m_2)+1$ labelled by the weight ${(m_1-m_2)}$. The $SU(2)$ weights must respect
\begin{equation}
n_1 + n_2 \ge m_1 \ge n_2 \ge m_2 \ge 0~.
\end{equation}
The corresponding $U(1)$ charges are $q = m_1 + m_2 - \tfrac{2}{3} (n_1 +2 n_2)$. 

\section{One-loop contribution for a free massive scalar}\label{scalar}

In this section, we consider the one-loop partition function of a minimally coupled free massive scalar, $\varphi(x^\mu)$, of mass squared $\mu^2$ on $\mathbb{C}P^2$ with Fubini-Study metric denoted by $g_{\mu\nu}$. The action is given by
\begin{equation}
S_E[\varphi] = \frac{1}{2} \int d^4 x \sqrt{g} \varphi \left( \Delta_{(0)} + \mu^2 \right) \varphi~.
\end{equation}
The $\mathbb{C}P^2$ partition function is given by
\begin{equation}\label{Zphi}
\mathcal{Z}^{(\varphi)}_{\mathbb{C}P^2} = \int \mathcal{D}\varphi e^{-S_E[\varphi]} = {\det}^{-1/2}  \left(\frac{\Delta_{(0)} + \mu^2}{\Lambda_{\text{u.v.}} }\right)~,
\end{equation}
where $\Lambda_{\text{u.v.}}$ is a reference scale, with units of inverse length squared, that can be taken to be the ultraviolet cutoff, needed to define the measure of the path integral. Specifically, we normalise our measure as
\begin{equation}
1 = \int \mathcal{D}\varphi \exp \left( {-\frac{\Lambda_{\text{u.v.}}}{2}\int d^4 x \sqrt{g} \varphi^2} \right)~.
\end{equation}
We now proceed to compute the functional determinant.

\subsection{Heat kernel method}

We will regularise the formal expression (\ref{Zphi}) by a heat kernel method, such that
\begin{equation}
\log \mathcal{Z}^{(\varphi)}_{\mathbb{C}P^2} = \sum_{n\in\mathbb{N}} \, d_n^{(0)}  \int_{\mathbb{R}^+} \frac{d\tau}{2\tau} e^{-\frac{\varepsilon^2}{4\tau}-\left(\lambda_n^{(0)} + \mu^2\right)\tau}~,
\end{equation}
where $\varepsilon^2 \equiv 2^2 e^{-2\gamma} \Lambda_{\text{u.v.}}^{-1}$ with $\gamma$ the Euler-Mascheroni constant. It is convenient to express part of the integrand as follows
\begin{equation}
e^{-\left(\lambda_n^{(0)} + \mu^2\right)\tau} = \frac{e^{-(\mu^2-1) \tau}}{\sqrt{4\pi\tau}} \int_{\mathbb{R}}  du \,  {e^{-\frac{u^2}{4\tau}}} \, e^{i (n+1)u }~.
\end{equation}
For the sake of simplicity, we have set $\Lambda=\tfrac{3}{2}$. One can reinstate $\Lambda$ by rescaling $\varepsilon^2 \to \tfrac{2\Lambda}{3}\varepsilon^2$ and $\mu^2 \to \tfrac{3}{2\Lambda}\mu^2$ in the final expression. 
%$\tau \to \tfrac{2\Lambda}{3}\tau$, 
We now note that
\begin{equation}
\sum_{n\in\mathbb{N}} d^{(0)}_n  e^{i (n+1)u } = \frac{e^{i u} \left(1+4 e^{i u}+e^{2 i u}\right)}{\left(1-e^{i u}\right)^4}~.
% \frac{1}{8} (\cos u+2) \csc ^4\left(\frac{u}{2}\right)~.
\end{equation}
Moreover, the $\tau$-integral yields
\begin{equation}
\int_{\mathbb{R}^+} \frac{d\tau}{2\tau \sqrt{4 \pi  \tau }} e^{-\frac{\varepsilon ^2}{4 \tau }- \frac{u^2}{4 \tau}} e^{ -\tau  \left(\mu ^2-1\right)}= \frac{e^{-\nu\sqrt{u^2+\varepsilon ^2}}}{2 \sqrt{u^2+\varepsilon ^2}}~, \quad\quad \nu\equiv \sqrt{\mu^2-1}~.
\end{equation}
Consequently, upon performing the $\tau$-integral, we obtain
\begin{equation}
\log \mathcal{Z}^{(\varphi)}_{\mathbb{C}P^2} =  \int_{\mathbb{R}+i\delta}  du\frac{e^{-\nu\sqrt{u^2+\varepsilon ^2}}}{2 \sqrt{u^2+\varepsilon ^2}} \frac{e^{i u} \left(1+4 e^{i u}+e^{2 i u}\right)}{\left(1-e^{i u}\right)^4}~,
\end{equation}
where $0<\delta< \varepsilon$. Following \cite{Anninos:2020hfj}, we define a new variable $u=i t$, and upon a contour deformation we find
\begin{equation}\label{logZphi}
\log \mathcal{Z}^{(\varphi)}_{\mathbb{C}P^2} =  \int_{\varepsilon}^\infty  \frac{dt}{2 \sqrt{t^2-\varepsilon ^2}} \frac{1+e^{-t}}{ 1-e^{-t} } \frac{1+4 e^{-t}+e^{-2t}}{1+e^{-t}} \frac{e^{- t+i \nu\sqrt{t^2-\varepsilon^2}}+e^{- t-i \nu\sqrt{t^2-\varepsilon^2}}}{\left(1-e^{-t}\right)^3}~.
%(1+2 e^{-t})^2
%1+4 e^{-t}+e^{-2t}
\end{equation}
Upon formally setting $\varepsilon=0$, we can perform a small-$t$ expansion of the integrand to find
\begin{equation}\label{smallt}
{I}_\nu(t) =  \frac{6}{t^5}-\frac{3 \nu ^2}{t^3}+\left({\frac{\nu ^4}{4}+\frac{1}{120}}\right)\frac{1}{t} + \ldots
\end{equation}
The two leading ultraviolet divergences are associated to the local counterterms. If we couple the theory to gravity they can be absorbed into the  cosmological constant, $\Lambda$, and the Newton constant, $G$. The $\tfrac{1}{t}$ divergence is a logarithmic divergence that often appears for even-dimensional quantum field theories. The coefficient of the logarithmic divergence can also be calculated through other means, and the general result for an Einstein metric satisfying $R_{\mu\nu} = \Lambda g_{\mu\nu}$ follows from equation (4.12) of \cite{Christensen:1978md} which, for a minimally coupled scalar field of mass squared $\mu^2$,  yields
\begin{eqnarray}\label{B4scalar} \nonumber
B^{{\mathbb{C}P^2}}_4 &=& \frac{1}{180(4\pi)^2}\left( 32\pi^2 \chi_{\mathbb{C}P^2} +\frac{R^2}{6} V_{\mathbb{C}P^2} \left(\frac{29}{2} - 1 \right) + V_{\mathbb{C}P^2}  \left(90 \mu^4 - 30 R \mu^2 \right)  \right) \\
 &=&  \frac{31}{120}+\frac{9 \mu ^4}{16 \Lambda ^2}-\frac{3 \mu ^2}{4 \Lambda }~.
\end{eqnarray}
Recalling that we have set $\Lambda=\tfrac{3}{2}$  we must take $\mu^2 \to \tfrac{3}{2 \Lambda}\mu^2$ for the $\tfrac{1}{t}$ coefficient in (\ref{smallt}) to indeed find agreement.  
\newline\newline
In appendix \ref{scalarZapp}, we re-derive (\ref{logZphi}) directly from a Green's function analysis, in the coincident point limit, using methods developed in  \cite{Bandaru:2024qvv}. Happily, we land on (\ref{frederik}), which yields precisely the same expression as (\ref{logZphi}) upon formally setting $\varepsilon$ to zero. This constitutes a non-trivial test of the scalar Laplacian eigenspectrum (\ref{scalarspec}).
%\begin{equation}
%b_4 = \frac{53 \Lambda ^2}{240 \pi ^2} V_{\mathcal{M}_4} +\frac{\chi_{\mathcal{M}_4}}{90}~.
%\end{equation}
%Evaluating the above expression for $\mathbb{C}P^2$ yields $b_4 = $.

\subsection{Evaluation of heat kernel integral} 

Although the expression (\ref{logZphi}) diverges in the limit $\varepsilon \to 0^+$, upon taking sufficient $\nu$ derivatives, the resulting expression becomes finite. Taking five $\nu$-derivatives, and integrating back up, we deduce that
\begin{multline}
\log \mathcal{Z}^{(\varphi)}_{\mathbb{C}P^2} = \frac{i}{2}  \left(\nu ^3\psi ^{(-1)}(1+i \nu )  +3 i \nu ^2 \psi ^{(-2)}(1+i \nu) -6 \nu  \psi ^{(-3)}(1+i \nu) \right. \\ \left.  -6 i \psi ^{(-4)}(1+i \nu)\right)  + (\nu \leftrightarrow -\nu) + \sum_{n=0}^4 \frac{\alpha_n}{n!} \nu^n~,
%3 (\psi ^{(-4)}(1+i \nu )+\psi ^{(-4)}(1-i \nu ))-\frac{1}{2} i \nu  (\nu  (-\nu  \text{log$\Gamma $}(i \nu +1)+\nu  \text{log$\Gamma $}(1-i \nu ) \\ -3 i (\psi ^{(-2)}(i \nu +1)+\psi ^{(-2)}(1-i \nu )))+6 \psi ^{(-3)}(i \nu +1)-6 \psi ^{(-3)}(1-i \nu )) 
%-\frac{1}{2} \nu ^3 (\psi(1+i \nu )+\psi(1-i \nu)) + \sum_{n=0}^4 \alpha_n \nu^n~,
\end{multline}
%\begin{equation}
%\frac{i}{2}  \left(\nu ^3\psi ^{(-1)}(1+i \nu )+3 i \nu ^2 \psi ^{(-2)}(1+i \nu)-6 \nu  \psi ^{(-3)}(1+i \nu)-6 i \psi ^{(-4)}(1+i \nu)\right)
%%-\frac{1}{2} i \left(\nu ^3 \text{log$\Gamma $}(1-i \nu )-3 i \nu ^2 \psi ^{(-2)}(1-i \nu )-6 \nu  \psi ^{(-3)}(1-i \nu )+6 i \psi ^{(-4)}(1-i \nu )\right)
%\end{equation}
where the coefficients $\alpha_n$ can diverge in the $\varepsilon \to 0^+$ limit. The function $\psi^{(n)}(z)$ is the polygamma function of order $n$. Of these divergences, one will be logarithmic in $\varepsilon$, and the coefficient is given by the $\sim t^{-1}$ term in (\ref{smallt}). Moreover, we can evaluate some of the $\nu$-derivatives of the heat kernel integral (\ref{logZphi}) at $\nu=0$. For instance, taking an odd number of $\nu$ derivatives and setting $\nu=0$, the integral vanishes trivially. Taking four $\nu$-derivatives and setting $\nu=0$ appears to yield 
\begin{equation}
\partial_\nu^4 \log \mathcal{Z}^{(\varphi)}_{\mathbb{C}P^2}|_{\nu=0}  = 3 + 6\log 2 + 6 \log\frac{1}{\varepsilon} + \ldots~,
\end{equation}
to good numerical accuracy. Similarly,
\begin{equation}
\partial_\nu^2 \log \mathcal{Z}^{(\varphi)}_{\mathbb{C}P^2}|_{\nu=0}  = \frac{1}{12} - \frac{2}{\varepsilon^2} + \ldots~,
\end{equation} 
to good numerical accuracy. Finally, 
\begin{equation}\label{063}
\log \mathcal{Z}^{(\varphi)}_{\mathbb{C}P^2}|_{\nu=0}  = \frac{4}{\varepsilon^4}  + \frac{1}{120} \log\frac{1}{\varepsilon}  + 0.0063 \ldots~,
\end{equation} 
to good numerical accuracy. The above results are sufficient to fully fix the heat kernel integral (\ref{logZphi}) up to terms that vanish in the $\varepsilon \to 0^+$ limit. 
\newline\newline
Although the above numerical methods are sufficient to get a good approximation for the heat kernel integral (\ref{logZphi}), we note that the method introduced and detailed in appendix C.1 of \cite{Anninos:2020hfj} provides a more powerful  alternative. Indeed, employing equation (C.19) of \cite{Anninos:2020hfj} to the case (\ref{logZphi}) at hand yields the exact expression
\begin{equation}\label{scalarZ}
\log \mathcal{Z}^{(\varphi)}_{\mathbb{C}P^2} =  f(\nu) - \frac{\nu^4}{3} + \left(\frac{\nu^4}{4}+ \frac{1}{120}  \right) \log\frac{2 e^{-\gamma}}{\varepsilon} -\frac{\nu^2}{\varepsilon^2} +\frac{4}{\varepsilon^4}~,
\end{equation}
where 
\begin{equation}
f(\nu) \equiv \frac{1}{2}\sum_\pm (\hat{\delta}\mp i \nu )^3 \zeta'(0,1\pm i \nu)~, 
\end{equation}
with $\hat{\delta}^n \zeta'(0,1\pm i \nu) \equiv \zeta'(-n,1\pm i \nu)$. The function $\zeta(z,\Delta)$ is the Hurwitz zeta function. The results are compatible with the numerical expressions, and complete them. For example, the approximate number $0.0063$ becomes the exact quantity $\zeta '(-3)-\frac{\gamma }{120}+\frac{\log 2}{120}$. One can use similar methods to evaluate the heat kernel integrals that appear in the later sections also.

\subsection{Comparison to $S^4$} 

It is interesting to compare our expression (\ref{logZphi}) to the analogous expression found in \cite{Anninos:2020hfj} for a minimally coupled free scalar of mass $\mu$ on a round four-sphere. The latter reads
\begin{equation}\label{ZS4phi}
\log \mathcal{Z}^{(\varphi)}_{S^4} = \int_{\varepsilon}^\infty  \frac{dt}{2 \sqrt{t^2-\varepsilon ^2}} \frac{1+e^{-t}}{ 1-e^{-t} }  \frac{e^{- \frac{3t}{2}+i \tilde{\nu}\sqrt{t^2-\varepsilon^2}}+e^{-\frac{3 t}{2}-i \tilde{\nu}\sqrt{t^2-\varepsilon^2}}}{\left(1-e^{-t}\right)^3}~.
\end{equation}
Here we have defined $\tilde{\nu} \equiv \sqrt{3\mu^2\Lambda^{-1} -\tfrac{9}{4}}$. Upon formally setting $\varepsilon=0$, the integrand of above expression contains the $SO(4,1)$ Harish-Chandra character for the principal series irreducible representation labelled by the quadratic Casimir $\mathcal{C}_{\tilde{\nu}} \equiv \Delta_{\tilde{\nu}}(3-\Delta_{\tilde{\nu}})$ with $\Delta_{\tilde{\nu}} =\tfrac{3}{2} + i \tilde{\nu}$. The Harish-Chandra character reads
\begin{equation}
\chi_{\Delta_{\tilde{\nu}}}(t) =  \frac{e^{-\Delta_{\tilde{\nu}} t }+e^{-(3-\Delta_{\tilde{\nu}})t}}{\left(1-e^{-t}\right)^3}~.
\end{equation}
It was further noted in \cite{Anninos:2020hfj} that one can express the partition function in a form that resembles that of a thermal ideal gas. In the formal $\varepsilon=0$ limit this reads
\begin{equation}\label{Zthermal}
\log \mathcal{Z}^{(\varphi)}_{S^4} = \int_{\mathbb{R}^+} d\omega \rho_{\tilde{\nu}}(\omega) \log \left(e^{\frac{\beta \omega}{2}}-e^{-\frac{\beta \omega}{2}} \right)^{-1}~, \quad\quad \beta \equiv 2\pi \sqrt{\tfrac{3}{\Lambda}}~.
\end{equation}
Here $\rho_{\tilde{\nu}}(\omega)$ is a density of states obtained by Fourier transforming $\chi_{\Delta}(t)$. When evaluating the heat kernel integral (\ref{ZS4phi}), Hurwitz zeta functions of the type $\zeta'(-n,\tfrac{3}{2} \pm i \tilde{\nu})$ with $n=0,1,2,3$ appear (as an example, see (C.27) with $g_s=1$ in \cite{Anninos:2020hfj}).
\newline\newline
We can also compare the leading heat kernel coefficients for a massive scalar on $\mathbb{C}P^2$ to the ones for a massive scalar on $S^4$. Reinstating the $\Lambda$ dependence in (C.13) of \cite{Anninos:2020hfj}, the leading heat kernel divergence is found to be $B^{S^4}_0 = \tfrac{4}{3} \left( \tfrac{3}{\Lambda} \right)^2 \varepsilon^{-4}$. For $\mathbb{C}P^2$, we can read off from (\ref{scalarZ}) that $B^{\mathbb{C}P^2}_0 = 4  \left( \tfrac{3}{2\Lambda} \right)^2 \varepsilon^{-4}$. Their ratio is then
\begin{equation}\label{B0r}
\frac{B^{S^4}_0}{B^{\mathbb{C}P^2}_0} = \frac{4}{3} = \frac{V_{S^4}}{V_{\mathbb{C}P^2}}~,
\end{equation}
as expected. Similarly, one finds that the ratio of the second heat kernel coefficients is similarly given by
\begin{equation}\label{B2r}
\frac{B^{S^4}_2}{B^{\mathbb{C}P^2}_2} = \frac{4}{3}~.
% = \frac{\int_{S^4} \sqrt{g} R}{\int_{\mathbb{C}P^2}  \sqrt{g} R}~.
\end{equation}
The ratio of the coefficients for the logarithmic divergence (\ref{B4scalar}) are no longer related to each other by a simple pre-factor.

\subsection{Non-perturbative features}

If the massive scalar field is weakly coupled to gravity with $\Lambda>0$, then in the saddle-point approximation the Euclidean gravitational path integral may contain contributions from the $S^4$ and $\mathbb{C}P^2$ saddles. These will be weighted by the respective on-shell actions, and the one-loop contributions. At one-loop
\begin{equation}
\mathcal{Z}_{\text{grav}} \approx e^{\mathcal{S}_{S^4}}  \mathcal{Z}^{(1)}_{S^4} \mathcal{Z}^{(\varphi)}_{S^4} + e^{\mathcal{S}_{\mathbb{C}P^2}} \mathcal{Z}^{(1)}_{\mathbb{C}P^2} \mathcal{Z}^{(\varphi)}_{\mathbb{C}P^2} + \ldots 
\end{equation}
We have denoted the one-loop contributions from the gravitational fluctuations by $\mathcal{Z}^{(1)}_{\mathcal{M}^4}$. The one-loop expression for $\mathcal{Z}^{(1)}_{S^4}$ was computed in \cite{Volkov:2000ih,Anninos:2020hfj}, and we consider $\mathcal{Z}^{(1)}_{\mathbb{C}P^2}$ in section \ref{Zgrav1sec}. Upon coupling to gravity, the ultraviolet divergences of the quantum field theory partition functions can be absorbed into the physical Newton constant, $G$, and cosmological constant, $\Lambda$. Once we pick a regularisation scheme for the quantum perturbation theory around one of the saddles, we must consistently implement this scheme about the other saddles. As such, the presence of multiple saddles permits us to consider a variety of scheme independent quantities by comparing the perturbative features around each saddle.
\newline\newline
We can view the contribution from $\mathbb{C}P^2$ as a non-perturbatively small correction to the leading $S^4$ contribution. The contribution from $\mathbb{C}P^2$ cannot be expressed as a $SO(4,1)$ Harish-Chandra character, so we interpret it as a small de Sitter symmetry breaking effect of size $e^{\mathcal{S}_{\mathbb{C}P^2}-\mathcal{S}_{S^4}} = e^{-\tfrac{3\pi}{4\Lambda G}}$. For the current value of the cosmological constant, this is an extraordinarily small number of the order $e^{-10^{122}}$.\footnote{In particular, it is exponentially smaller than the putative effects discussed in \cite{Tsamis:1996qq,Polyakov:2007mm}. It would be interesting to understand if it could compete with other non-perturbative instabilities, for instance when de Sitter space is realised as a metastable vacuum in string theory.} 
If we also add the contribution from higher-derivative terms, such as the Gauss-Bonnet term or the Weyl squared term, the suppression factor will also depend on the value of their respective couplings. We expect higher-derivative effects to be suppressed by some high energy scale in our present cosmology. Perhaps during primordial inflation non-spherical topologies  give rise to more enhanced contributions, particularly if they contribute to Lorentzian processes at sufficiently late-times. It is interesting to note, further, that in higher dimensions there exist saddles that break the de Sitter isometries \cite{boyer2005einstein,bohm1998inhomogeneous} on the topological sphere itself. 
%We can introduced a modified character
%\begin{equation}
%\chi_{\text{n.p.}} \equiv \chi_{\Delta_{\tilde{\nu}}}(t)  + e^{\mathcal{S}_{\mathbb{C}P^2}-{\mathcal{S}_{S^4}}} 
%\end{equation}

\section{One-loop contribution for the Maxwell field}\label{maxwell}

In this section, we consider the partition function of the Maxwell field on $\mathbb{C}P^2$. The theory consists of an Abelian gauge field $A_{\mu}(x^\mu)$ subject to the gauge transformation $A_\mu \to A_\mu +  \partial_\mu \omega$, where $\omega\sim\omega +\tfrac{2\pi}{\bold{e}}$ is a compact scalar of size $\tfrac{2\pi}{\bold{e}}$ where $\bold{e}$ is the dimensionless coupling of the theory. The computation serves as a check for the transverse vector spectra reported in \cite{Warner:1982fs,Pope:1980ub,boucetta2010spectra}, since we can compare against the heat kernel coefficients of \cite{Christensen:1978md,Vassilevich:1993yt}.   In appendix \ref{intmax} we provide explicit expressions for the various heat kernel integrals that appear. Gauge fields on $\mathbb{C}P^2$ have been considered in \cite{Karabali:2021gae,Karabali:2022qcr}. Finally, it was pointed out in \cite{Gibbons:1978zy} that we can view the Fubini-Study metric on $\mathbb{C}P^2$ as a solution to the Einstein-Maxwell equations with $\Lambda>0$, provided the electromagnetic field-strength $F_{\mu\nu}$ is proportional to the K\"ahler form of $\mathbb{C}P^2$.

\subsection{Path integral for the Maxwell field}

We take the action of our theory to be canonically normalised, such that
\begin{equation}
S_{U(1)} = \frac{1}{4 } \int d^4 x\sqrt{g} \left(\nabla_\mu A_\nu - \nabla_\nu A_\mu\right)\left(\nabla^\mu A^\nu - \nabla^\nu A^\mu\right)~.
\end{equation}
The path integral of a $U(1)$ gauge field $A_{\mu}$ on $\mathbb{C}P^2$, with gauge coupling $\bold{e}$, takes a form that is essentially identical to that over a four-sphere, and we can simply adapt the treatment of \cite{Donnelly:2013tia,Giombi:2015haa,Law:2020cpj} to our case. Formally, the path integral is given by
\begin{equation}\label{ZU1}
\mathcal{Z}_{U(1)} =  \frac{1}{\text{vol}_{U(1)}} \int \mathcal{D}A_\mu e^{-S_{U(1)}}~.
\end{equation}
We normalise our local measure over the space of gauge fields as
\begin{equation}
1 = \int \mathcal{D}A_{\mu} \exp\left( -\frac{\Lambda_{\text{u.v.}}}{2} \int d^4 x \sqrt{g}  A_{\mu}  A^{\mu} \right)~.
% \Lambda_{\text{u.v.}} 
\end{equation}
The locally defined metric for the gauge parameter $\omega(x^\mu)$ generating the $U(1)$ gauge group is given by
\begin{equation}
ds^2_{U(1)} = \frac{\Lambda_{\text{u.v.}}^2}{2\pi} \int d^4 x \sqrt{g} \delta\omega(x^\mu)^2~.
\end{equation}
%For the gauge parameter $\omega(x^\mu)$, we take the measure to be
%\begin{equation}
%1 = \int \mathcal{D}\delta\omega(x^\mu) \exp\left( - \Lambda_{\text{u.v.}}^2 \int d^4 x \sqrt{g} \delta\omega(x^\mu)^2 \right)~.
% \Lambda_{\text{u.v.}} 
%\end{equation}
%The locally defined volume of the gauge group is given by $\text{vol}_{U(1)} \equiv \Lambda^2_{\text{u.v.}} \int \mathcal{D}\omega$. 
As such, the full expression (\ref{ZU1}) is itself manifestly local. 
\newline\newline
To compute the path integral, we can decompose the gauge field as $A_\mu = A^T_\mu + \partial_\mu \varphi$, where $\nabla^\mu A^T_\mu = 0$ and $\varphi$ is a scalar function with the constant part removed. The path integral over  $\varphi$ cancels most of the volume $\text{vol}_{U(1)}$, except for the volume stemming from the constant $\omega$ mode, namely  
\begin{equation}
\text{vol}_{U(1)_c} = \frac{2\pi}{\bold{e}} \,   \sqrt{\frac{\Lambda^2_{\text{u.v.}}}{2\pi} \, V_{\mathbb{C}P^2}}~.
\end{equation}
We must also include the appropriate Jacobian factor that arises from the decomposition of $A_\mu$. All in all, the path integral thus becomes
\begin{equation}\label{ZU1f}
\mathcal{Z}^{U(1)}_{\mathbb{C}P^2}= \frac{\bold{e}}{\sqrt{2\pi \, \Lambda^2_{\text{u.v.}}  V_{\mathbb{C}P^2}}} \, \frac{ {\det}'^{\frac{1}{2}} \frac{\Delta_{(0)}}{\Lambda_{\text{u.v.}} }  }{  {\det}^{\frac{1}{2}} \frac{\Delta_{(1)}}{\Lambda_{\text{u.v.}}}}~,
\end{equation}
where the spectrum of the Hodge-De Rham operator $\Delta_{(1)}$ acting on transverse vector fields was given in (\ref{vecspec}) and (\ref{vecspec2}). The prime indicates we are omitting the zero mode of the scalar Laplace operator. We can follow the same steps as for the scalar field computation to express the functional determinants in terms of heat kernel integrals.

\subsection{Scalar determinant contribution}

For the scalar determinant appearing in (\ref{ZU1f}), we have to compute the heat kernel regularised logarithm of the functional determinant with the zero mode removed. Upon setting $\Lambda = \tfrac{3}{2}$, this gives rise to the heat kernel integral
\begin{equation}\label{massless}
\mathcal{I}_0 =    \int_{\varepsilon}^\infty  \frac{dt}{2 \sqrt{t^2-\varepsilon ^2}} \frac{4-e^{-t}-5 e^t+8 e^{2 t}}{(1-e^{t})^4}  \times \left(e^{\sqrt{t^2-\varepsilon^2}} + e^{-\sqrt{t^2-\varepsilon^2}} \right)~.
%-  \int_{\varepsilon}^\infty  \frac{dt}{2 \sqrt{t^2-\varepsilon ^2}} \frac{e^{3t}\left(5+e^{2t} -12 \cosh t + 4 \sinh t  \right)}{(1-e^{t})^4} \times \left(e^{\sqrt{t^2-\varepsilon^2}} + e^{-\sqrt{t^2-\varepsilon^2}} \right)~.
\end{equation}
From which we can extract, upon formally setting $\varepsilon=0$ in the integrand, the $\tfrac{1}{t}$ coefficient to be
\begin{equation}
\mathcal{B}_0 = \left(\frac{31}{120}-1 \right)  = -\frac{89}{120}~.
\end{equation}
The subtraction of $1$ comes from the fact that we have omitted the zero mode. As for the scalar heat kernel integral,  we can use the method presented in appendix C.1 of \cite{Anninos:2020hfj} to compute $\mathcal{I}_0$, as well as the heat kernel integrals that appear below, at small $\varepsilon$. The resulting expression for the heat kernel integral $\mathcal{I}_0$ is provided in \ref{I0}. 

\subsection{Transverse vector determinant contribution}

As for the scalar determinant we can perform a heat kernel analysis for the transverse vector determinant. Here there are no zero modes, so we can proceed with the whole spectra. For the first collection of eigenvalues (\ref{vecspec}), the calculation is identical to that of the massless scalar determinant, again with the zero mode omitted. Upon setting $\Lambda = \tfrac{3}{2}$, the resulting heat kernel integral is identical to the scalar case, and is again given by
\begin{equation}\label{vector1}
\mathcal{I}_{11} =    \int_{\varepsilon}^\infty  \frac{dt}{2 \sqrt{t^2-\varepsilon ^2}} \frac{4-e^{-t}-5 e^t+8 e^{2 t}}{(1-e^{t})^4}  \times \left(e^{\sqrt{t^2-\varepsilon^2}} + e^{-\sqrt{t^2-\varepsilon^2}} \right)~.
%-  \int_{\varepsilon}^\infty  \frac{dt}{2 \sqrt{t^2-\varepsilon ^2}} \frac{e^{3t}\left(5+e^{2t} -12 \cosh t + 4 \sinh t  \right)}{(1-e^{t})^4} \times \left(e^{\sqrt{t^2-\varepsilon^2}} + e^{-\sqrt{t^2-\varepsilon^2}} \right)~.
\end{equation}
From the above we can extract, upon formally setting $\varepsilon=0$, the $\tfrac{1}{t}$ coefficient of the small-$t$ expansion in the integrand to be
\begin{equation}
\mathcal{B}_{11} =\left(\frac{31}{120}-1 \right)= -\frac{89}{120}~.
\end{equation}
For the second set of eigenvalues  (\ref{vecspec2}), one instead finds the heat kernel integral
\begin{equation}\label{vector2}
\mathcal{I}_{12} = \int_{\varepsilon}^\infty  \frac{dt}{\sqrt{t^2-\varepsilon ^2}} \frac{e^{\frac{t}{2}}   (9 \sinh t+11 \cosh t-5)}{(1-e^{t})^4}  \times \left(e^{\frac{1}{2}\sqrt{t^2-\varepsilon^2}} + e^{-\frac{1}{2}\sqrt{t^2-\varepsilon^2}} \right)~.
%\cosh \tfrac{1}{2}\sqrt{t^2-\varepsilon^2}~.
%  \int_{\varepsilon}^\infty  \frac{dt}{\sqrt{t^2-\varepsilon ^2}} \frac{e^{\tfrac{7}{2}t}\left(-5+11 \cosh t -9 \sinh t  \right)}{(1-e^{t})^4}  \times \left(e^{\tfrac{1}{2}\sqrt{t^2-\varepsilon^2}} + e^{-\tfrac{1}{2}\sqrt{t^2-\varepsilon^2}} \right) 
%\cosh \tfrac{1}{2}\sqrt{t^2-\varepsilon^2}~.
\end{equation}
Upon formally setting $\varepsilon=0$, the $\tfrac{1}{t}$ coefficient of the integrand expanded at small-$t$ is given by
\begin{equation}
\mathcal{B}_{12} = \frac{19}{15}~.
\end{equation} 
The heat kernel integral $\mathcal{I}_{12}$ is evaluated in appendix \ref{intmax}, and the resulting expression is provided in (\ref{I1}). 
\newline\newline
Putting it all together, the heat kernel regularised expression for the Maxwell field on $\mathbb{C}P^2$ is given by
%\begin{multline}
%\log \mathcal{Z}^{U(1)}_{\mathbb{C}P^2}=  \log \bold{e}+ \frac{8}{\varepsilon ^4} \left(\frac{3}{2\Lambda} \right)^2-\frac{4}{\varepsilon^2}\left(\frac{3}{2\Lambda} \right)+\frac{11}{15}  \log \left(\sqrt{\frac{2\Lambda}{3}}  \frac{\varepsilon e^\gamma }{2} \right)  \\  +\frac{13}{48} - \log 6 \pi^{\frac{3}{2}} + 3 \log A+2 \zeta'(-3)~.
%\end{multline}
\begin{multline}
\log \mathcal{Z}^{U(1)}_{\mathbb{C}P^2}=  \log \bold{e}+ \frac{8}{\varepsilon ^4} \left(\frac{3}{2\Lambda} \right)^2-\frac{4}{\varepsilon^2}\left(\frac{3}{2\Lambda} \right)+\frac{11}{15}  \log \left(\sqrt{\frac{2\Lambda}{3}}  \frac{\varepsilon e^\gamma }{2} \right)  \\  +\frac{13}{48} - \log 4 \pi^{\frac{3}{2}} + 3 \log A+2 \zeta'(-3)~.
\end{multline}
If we recall that the Glaisher constant obeys the identity $\log A = \tfrac{1}{12} - \zeta'(-1)$, we can also replace $A$ in the above expression with a derivative of the Riemann zeta function.

\subsection{Comparing to the heat kernel coefficient}

As for the scalar field, the logarithmic coefficient can also be calculated through other means, and the general result for an Einstein metric satisfying $R_{\mu\nu} = \Lambda g_{\mu\nu}$ follows from equation (6.16) of \cite{Christensen:1979iy} which, for a $U(1)$ gauge field yields
\begin{equation}\label{B4m}
B^{{\mathbb{C}P^2}}_4 =  \frac{1}{180(4\pi)^2}\left(-13 \times 32\pi^2 \chi_{\mathbb{C}P^2} -48 \Lambda^2 V_{\mathbb{C}P^2}   \right) = -\frac{11}{15}~.
\end{equation}
If we sum the coefficients of the various $\tfrac{1}{t}$ coefficients arising in the path integral computation we find
\begin{equation}
\left( - \frac{89}{120} + \frac{19}{15}  \right) - \left( \frac{31}{120} -1 \right)- 2 = -\frac{11}{15}~.
\end{equation}
We note that, as detailed in appendix G of \cite{Anninos:2020hfj},  the final $-2$ stems from a proper treatment of the path integration measure.  It is precisely the $\Lambda_{\text{u.v.}}^{-1} \sim \varepsilon^2$ prefactor appearing in (\ref{ZU1f}). A similar treatment can be employed for the Maxwell theory on a round $S^4$, see for example  \cite{Giombi:2015haa,Anninos:2020hfj,RiosFukelman:2023mgq}, which yields a coefficient
\begin{equation}
B^{{S^4}}_4 = \frac{19}{30}-\left(\frac{29}{90}-1\right)-2 =-\frac{31}{45}~,
\end{equation}
where we have expressed the coefficient in terms of the transverse vector, massless scalar, and zero mode split. This indeed agrees with the $S^4$ version of (\ref{B4m}). Similarly, we can compute the ratios of the leading heat kernel coefficients and find the ratio $\tfrac{4}{3}$, as in (\ref{B0r}) and (\ref{B2r}). Our heat kernel coefficients for $\mathbb{C}P^2$ also agree with those computed in \cite{Vassilevich:1993yt} up to zero mode omissions.

\section{One-loop contribution for the linearised graviton}\label{Zgrav1sec}

In this section we consider the gravitational path integral for $\Lambda>0$ about the $\mathbb{C}P^2$ saddle. We begin with the Einsten-Hilbert action with a cosmological term
\begin{equation}
S_E = \frac{1}{16\pi G} \int d^4x \sqrt{g} \left(-R+2\Lambda\right)~.
\end{equation}
We consider small perturbations $h_{\mu\nu}$ about the Fubini-Study solution in (\ref{FS}). We are interested in computing the one-loop path inegral
\begin{equation}
\mathcal{Z}^{(1)}_{\mathbb{C}P^2} = \frac{1}{\text{vol}_{\text{diff}}}\int \mathcal{D} h_{\mu\nu} e^{-S^{(2)}_E[h_{\mu\nu}]}~,
\end{equation}
where $S^{(2)}_E$ is the quadratic action for the linearised metric perturbation. To do so we will proceed by defining the measure and volume of the diffeomorphism group in order to get an expression in terms of functional determinants. In appendix \ref{intgrav} we provide explicit expressions for the various heat kernel integrals that appear.

\subsection{Linearised action and measure}

Following the conventions of \cite{Law:2020cpj}, we decompose the linearised perturbation as 
\begin{equation}\label{decom}
h_{\mu\nu} = \frak{h}_{\mu\nu} + \frac{1}{\sqrt{2}} \left( \nabla_\mu \xi_\nu + \nabla_\nu \xi_\mu \right) + \frac{g_{\mu\nu}}{2} \tilde{h}~,
\end{equation}
where $\frak{h}_{\mu\nu}$ is transverse and traceless with respect to the background metric $g_{\mu\nu}$. We require that the vector fields $\xi_\mu$ are orthogonal to the Killing vector fields of $\mathbb{C}P^2$, as these do not contribute to $h_{\mu\nu}$. Unlike the four-sphere, the Fubini-Study metric admits no non-isometric conformal Killing vector fields \cite{yano1959einstein}. 
\newline\newline
The canonically normalised linearised action splits into a contribution from the transverse-traceless part
\begin{equation}\label{Stt}
S_E^{(TT)} = \frac{1}{2} \int d^4 x \sqrt{g} \,\left( - \frak{h}^{\mu\nu} \nabla_\alpha \nabla^\alpha   \frak{h}_{\mu\nu} - 2 \frak{h}^{\mu\nu} R_{\mu\nu\rho\sigma} \frak{h}^{\rho\sigma} \right)~,
%\frac{1}{64\pi G}
\end{equation}
and the linearised action for the trace 
\begin{equation}
S_E^{(T)} = - \frac{3}{4} \int d^4 x \sqrt{g}   \tilde{h}  \left( \Delta_{(0)} - \frac{4\Lambda}{3}\right) \tilde{h}~. 
% \frac{1}{32\pi G}
\end{equation}
We note the kinetic term in the action $S_E^{(T)}$ has the wrong sign. This is the linearised version of the conformal mode problem, and we will treat it along the lines of \cite{Gibbons:1978ac}. Namely we will rotate the contour $\tilde{h} \to i \tilde{h}$ and rotate back any residual negative modes.\footnote{It was recently argued \cite{Ivo:2025yek} that a more sophisticated continuation prescription for the overall phase is required to ensure gauge invariance of the path integral. We will not keep track of the overall sign of the phase to this degree of precision throughout this section.} Due to the absence of non-isometric conformal Killing vectors, the only negative mode is in fact the constant mode of $\tilde{h}$. This also follows from a theorem of Lichnerowicz and Obata  \cite{lichnerowicz1958geometrie,obata1971conjectures}. Thus, the phase of the one-loop partition function is $\pm i$.  
\newline\newline
So far we have the contribution from the transverse-traceless and trace modes to the one-loop partition function. We must also compute the Jacobian that stems from the decomposition (\ref{decom}). This was done, for example, in \cite{Law:2020cpj} (see also \cite{Volkov:2000ih,Vassilevich:1993yt}), and here we review briefly the computation. The only modification,  as compared to the case of $S^4$, is the absence of non-isometric conformal Killing vector fields. We normalise the local measure of the path-integral as
\begin{equation}
1 = \int \mathcal{D}h_{\mu\nu} \exp\left( - \frac{\Lambda_{\text{u.v.}}}{2} \int d^4 x \sqrt{g} h_{\mu\nu} h^{\mu\nu} \right)~.
%, \quad\quad \bold{g}^2  \equiv 32\pi G~.
\end{equation}
%and similarly for other fields. 
Given our measure, we can compute the Jacobian factor upon decomposing the linearised perturbation. In addition to the decomposition (\ref{decom}), it is further convenient to decompose the vector fields into their transverse and longitudinal parts, namely
\begin{equation}
\xi_\mu = \eta_\mu + \partial_\mu \sigma~, \quad\quad \nabla^\mu \eta_\mu = 0~.
\end{equation}
Here $\sigma$ is a non-constant scalar, and $\eta_\mu$ is a transverse vector field which is not a Killing vector field of $\mathbb{C}P^2$. We do not need to worry about non-isometric conformal Killing vector fields for $\mathbb{C}P^2$ as they do not exist. A fairly straightforward calculation yields the Jacobian determinant factor
\begin{equation}
J = \frac{W_\sigma^+}{Y_\eta^T Y^+_\sigma}~,
\end{equation}
where we have defined
\begin{eqnarray}
W_\sigma^+ &\equiv& \int \mathcal{D}'\sigma \, e^{- \frac{\Lambda^2_{\text{u.v.}} }{2} \int d^4 x \sqrt{g} \sigma \Delta_{(0)} \sigma}~, \\ %\tfrac{1}{2}
Y_\sigma^+ &\equiv& \int \mathcal{D}'\sigma \, e^{-  \frac{3 }{4} \Lambda_{\text{u.v.}}  \int d^4 x \sqrt{g} \sigma \Delta_{(0)} \left( \Delta_{(0)} - \frac{4\Lambda}{3}\right) \sigma}~, \\ %\tfrac{1}{2}
Y_\eta^T &\equiv& \int \mathcal{D}'\eta \, e^{-\frac{\Lambda_{\text{u.v.}}}{2}   \int d^4 x \sqrt{g} \, \eta_\mu  \left( \Delta_{(1)} -2\Lambda\right) \eta^{\mu} }~. %\tfrac{1}{2}
%-\nabla_{(1)}^2 -\Lambda
\end{eqnarray}
Here, the prime in $\mathcal{D}'$ indicates that we are either omitting the Killing vector fields or the constant mode of $\sigma$. The factor $W_\sigma^+$ stems from the Jacobian due to decomposing $\xi_\mu$ into $\eta_\mu$ and $\sigma$. We note that 
% ({\color{blue} recheck, it comes from the zero mode of $\tilde{h}$})
%, and $\nabla_{(1)}^2$ is the Laplacian acting on transverse vector fields
\begin{equation}
\frac{W_\sigma^+  \mathcal{Z}_{\tilde{h}}}{ Y^+_\sigma}  = \pm  \frac{i}{2^{3/4} \times \sqrt{3\pi}} \left( V_{\mathbb{C}P^2} {\Lambda^2_{\text{u.v.}}} \right)^{\frac{1}{4}}~.
%2\times 
% \frac{i}{2 \times 2^{\tfrac{3}{4}}} \sqrt{\frac{\pi }{3}} 
%= \pm i xxx \sqrt{\frac{\Lambda_{\text{u.v.}}}{\Lambda}} =
%{\det}^{-\tfrac{1}{2}}\left(-\frac{\tfrac{3}{4} \left( \Delta_{(0)} -\tfrac{4\Lambda}{3}\right)}{\Lambda_{\text{u.v.}}}\right) \times  {\det}^{-\tfrac{1}{2}}\left(-\frac{\tfrac{3}{4} \left( \Delta_{(0)} -\tfrac{4\Lambda}{3}\right)}{\Lambda_{\text{u.v.}}}\right)
\end{equation}
As a final note, we could have also considered the approach of \cite{Christensen:1979iy,Polchinski:1988ua} and taken a gauge-fixing procedure. The gauge-fixing choice employed in \cite{Christensen:1979iy,Polchinski:1988ua} is the de Donder gauge whereby $\nabla^\mu h_{\mu\nu} - \tfrac{1}{2} g_{\mu\nu} h = 0$. In this gauge, the quadratic action governing the trace part of the metric perturbation appears with an operator $\Delta_{0}-2\Lambda$ which has eight zero modes from the $n=1$ sector of (\ref{scalarspec}). Something similar holds for the $S^2\times S^2$ saddle analysed in \cite{Volkov:2000ih}. These zero modes are an artefact of the gauge choice, although amusingly $i^8=1$.
 
\subsection{Group volume factor} 
 
We must also compute the contribution from the volume of the diffeomorphism group. Specifically we have a term in the gravitational  path integral
\begin{equation}\label{vold}
\frac{1}{\text{vol}_{\text{diff}}} \times \int \mathcal{D}'\xi \equiv \frac{1}{\text{vol}_{\text{diff}_c}}~,
\end{equation}
where the locally defined volume of the diffeomorphism group  stems from the metric
\begin{equation}\label{voldiff}
%\text{vol}_{\text{diff}} 
ds^2_\xi \equiv  \frac{\Lambda_{\text{u.v.}}^2}{2\pi} \int d^4 x \sqrt{g} \, \delta \xi^\mu \delta \xi_\mu~.
\end{equation}
Most of $\text{vol}_{\text{diff}}$ cancels against the $\int \mathcal{D}'\xi$ in the numerator of (\ref{vold}). What is left is a residual volume $\text{vol}_{\text{diff}_c}$ due to the fact that the path integral over the vector fields in the numerator did not include the Killing vector fields of $\mathbb{C}P^2$. We can express the Killing vector fields of $\mathbb{C}P^2$ in terms of an ambient space picture. Consider the hypersurface  $\delta_{I \bar{J}} Z^I \bar{Z}^{\bar{J}} = \tfrac{6}{\Lambda}$ with $I=1,2,3$ inside of $\mathbb{R}^6$, corresponding to an $S^5$. The Fubini-Study metric on $\mathbb{C}P^2$ is the space of orbits under the action $Z^I \sim e^{i\alpha} Z^I$. One representation of the Killing algebra generating the $SU(3)$ isometry group, can be written in terms of the embedding space coordinates as %({\color{blue}careful with complex coords, remember $\sqrt{2}$ in diffeo})
\begin{equation}\label{su3}
M^a = - \frac{1}{\sqrt{16\pi G}}  Z^I {[\lambda^a]_I}^J \,  {\partial}_J~,
%  \frac{1}{\sqrt{2}}
%_{I\bar{J}} 
%-  \bar{Z}^J {\partial}_I  \right)~.
\end{equation} 
where the $\lambda^a$, with $a=1,2,\ldots,8$, are Gell-Mann matrices normalised as $\text{Tr} \lambda^a \lambda^b = 2 \delta_{ab}$. As a concrete example, in terms of the coordinate system (\ref{taubNUT}), the Killing vector fields of the $su(2)$ subalgebra of $su(3)$ are given by
\begin{eqnarray}
  \xi^{\mu}_{(1)} \partial_\mu &=& \frac{2}{\sqrt{16\pi G}} \left(  \text{csc}\theta \sin\varphi \partial_\psi  + \cos\varphi \partial_\theta - \sin\varphi \cot \theta \partial_\varphi \right)~, \\
 \xi^{\mu}_{(2)} \partial_\mu &=& \frac{2}{\sqrt{16\pi G}}  \left( \text{csc}\theta \cos\varphi \partial_\psi  - \sin\varphi \partial_\theta - \cos\varphi \cot \theta \partial_\varphi \right)~, \\ 
  \xi^{\mu}_{(3)} \partial_\mu &=&  \frac{2}{\sqrt{16\pi G}} \partial_\varphi~. 
\end{eqnarray}
We have normalised our Killing vector fields in a way compatible with our rescaling of the metric fluctuation leading to the canonically normalised linearised action (\ref{Stt}).
%What we must do is compute this residual volume, as normalised by the path integral itself. The volume of the diffeomorphism group is itself obtained from a local measure over the space of vector fields, which we take to be 
%\begin{equation}\label{alphanorm}
%1 = \int \mathcal{D}\alpha_\mu \, e^{-\frac{1}{2\bold{g}^2} \int d^4 x \sqrt{g} \alpha_\mu \alpha^{\mu}}~.
%\end{equation}
The residual volume is the volume of the isometry group of $\mathbb{C}P^2$ as normalised with respect to (\ref{voldiff}) (see appendix G.4 of \cite{Anninos:2020hfj} for the analogous treatment of $S^4$). The invariant bilinear form on the Killing vector field algebra, with our path integral normalisation, is then 
%\cite{Joung:2013nma} 
%({\color{blue} check $1/2\pi$ remember its not volume of $S^5$})
\begin{equation}
%\langle M^1 | M^1  \rangle_{\text{PI}} = 
\frac{\Lambda_{\text{u.v.}}^2}{2\pi}  \int_{\mathbb{C}P^2} d^4 x \sqrt{g} \, \xi^\mu \xi_\mu = \frac{1}{2\pi} \frac{\Lambda_{\text{u.v.}}^2}{16\pi G} \times \frac{3}{\Lambda}  \times V_{\mathbb{C}P^2}~.
% {\color{blue} \frac{2}{3} \times \frac{6}{\Lambda} } \times  V_{\mathbb{C}P^2}~.
%_{1\bar{2}}
%_{\bar{1}{2}}
%\frac{1}{2\pi}
\end{equation}
%\begin{equation}
%\langle M^1 | M^1  \rangle_{\text{PI}} = \frac{1}{2\pi} \frac{\Lambda_{\text{u.v.}}^2}{16\pi G}  \int_{\mathbb{C}P^2} \left(Z^1 \bar{Z}^1 + Z^2 \bar{Z}^2  \right) = \frac{1}{2\pi} \frac{\Lambda_{\text{u.v.}}^2}{16\pi G} \times  {\color{blue} \frac{2}{3} \times \frac{6}{\Lambda} } \times  V_{\mathbb{C}P^2}~.
%%_{1\bar{2}}
%%_{\bar{1}{2}}
%%\frac{1}{2\pi}
%\end{equation}
The resulting residual group volume is thus
\begin{equation}
\text{vol}_{\text{cdiff}} =  \left(\frac{1}{2\pi}\right)^4 \times \left( \frac{1}{16\pi G \Lambda} \right)^4 \times {3}^4 \times  \left( \Lambda^2_{\text{u.v.}}  V_{\mathbb{C}P^2} \right)^4   \, \text{vol}_{\text{can}}~.
\end{equation}
%\begin{equation}
%\text{vol}_{\text{cdiff}} =  \left(\frac{1}{2\pi}\right)^4 \times \left( \frac{1}{16\pi G \Lambda} \right)^4 \times \left( {\color{blue}\frac{2}{3} \times {6} } \right)^4 \times  \left( \Lambda^2_{\text{u.v.}}  V_{\mathbb{C}P^2} \right)^4   \, \text{vol}_{\text{can}}~.
%\end{equation}
The volume $\text{vol}_{\text{can}} = \tfrac{\sqrt{3}\pi^5}{{3}}$ is the canonical group theoretic volume of $\tfrac{SU(3)}{\mathbb{Z}_3}$ (see for example \cite{boya2003volumes}).
%, where the generators of $SU(3)$ are normalised in a standard way such that $\langle M^{1} | M^{1} \rangle_{c} =1$.
% (i.e. (\ref{su3}) without the $\sqrt{16\pi G}$ factor).  
%\int_{\mathbb{C}P^2} M\cdot M \,  = 18\pi^2 \Lambda^{-2}
%Thus, the residual gauge group volume is xxx
%\subsection{Final one-loop expression}

\subsection{Expression for one-loop path integral}

Putting it all together, the one-loop  gravitational partition function reads
\begin{multline}\label{Zgrav1}
\mathcal{Z}^{(1)}_{\mathbb{C}P^2} =  {\mathcal{A}} \times \frac{\pm i}{\text{vol}_{\text{can}}}   \times \mathcal{S}_{\mathbb{C}P^2}^{-4} \times (V_{\mathbb{C}P^2} \Lambda^{2}_{\text{u.v.}})^{-4+\tfrac{1}{4}}   \\ \times {{{{\det}'^{\tfrac{1}{2}} \left( \frac{\Delta_{(1)} - 2\Lambda}{\Lambda_{\text{u.v.}}}\right)}} \times {{{\det}^{-\tfrac{1}{2}} \left( \frac{\Delta_{(2)} - 2\Lambda}{\Lambda_{\text{u.v.}}} \right)}}}~.
\end{multline}
We recall that $ \mathcal{S}_{\mathbb{C}P^2} =  \frac{\Lambda}{8\pi G} V_{\mathbb{C}P^2}$ and $V_{\mathbb{C}P^2} = \tfrac{18\pi^2}{\Lambda^2}$. In the above, $\mathcal{A}$ is given by
\begin{equation}\label{AZ}
\mathcal{A} = (2\pi)^4 \times  \frac{(6\pi)^8}{2^{\frac{3}{4}} \times \sqrt{3\pi} \times 3^{4}} ~.
%  =(2\pi)^4 \times 2^{-\frac{3}{4}} \, {(3\pi)^{\frac{15}{2}}}~.
%\left( {\color{blue}\frac{2}{3} \times {6} } \right)^{-4}
% = 6561 \pi^8~.
\end{equation}
Before proceeding to compute the determinants, we briefly compare to the case of $S^4$.
\newline\newline
\textbf{Comparison to $S^4$.} The above expression bears a clear resemblance to the analogous expression for $S^4$ \cite{Law:2020cpj}. The main differences are that in the above expression, we only have a phase from the constant mode of $\tilde{h}$, whereas in the $S^4$ expression the phase is $i^{(1+5)}$ as it receives contributions from the five non-isometric conformal Killing vector fields of $S^4$. Also the term $\mathcal{S}^{-4}_{\mathbb{C}P^2}$ becomes  $\mathcal{S}^{-5}_{S^4}$ reflecting the difference in dimension of the respective isometry groups of $S^4$ and $\mathbb{C}P^2$.

%\left(\frac{SU(3)}{\mathbb{Z}_3}

\subsection{Heat kernel integrals}

We now proceed to examine the functional determinants given the spectra in section \ref{spectra}. We begin with the determinant of the transverse vector fields and then move on to the determinant for the transverse and traceless tensor.

\subsubsection*{Vector determinant}

For the vector fields, the appropriate spectra are a shifted version of (\ref{vecspec}) and (\ref{vecspec2}) such that the eight Killing vectors have vanishing eigenvalue. Explicitly, the shifted spectrum is given by
\begin{eqnarray} 
\tilde{\lambda}^{(1)}_n &=& \frac{2\Lambda}{3}n(n+2)-2\Lambda~, \quad\quad \tilde{d}^{(11)}_n = (n+1)^3~,  \quad\quad n -1 \in \mathbb{Z}^+~, \\
\tilde{\gamma}^{(1)}_n &=& \frac{2\Lambda}{3}(n+1)(n+2)-2\Lambda~, \quad\quad \tilde{d}^{(12)}_n = n(n+3)(2n+3)~, \quad\quad n \in \mathbb{Z}^+~. \label{spec12}
\end{eqnarray}
Following the same procedure as for the scalar determinant, and setting $\Lambda=\tfrac{3}{2}$, we find the following contributions. From the first set of eigenvalues %({\color{blue}I must re-check})
\begin{equation}\label{vec1Z}
\tilde{\mathcal{I}}_{11} = \int_{\varepsilon}^\infty  \frac{dt}{\sqrt{t^2-\varepsilon^2}} \frac{-8 e^{-2 t}+31 e^{-t}+27 e^t-44}{\left(e^t-1\right)^4 }\times \cosh \left(2 \sqrt{t^2-\varepsilon ^2} \right)~.
% \int_{\varepsilon}^\infty  \frac{dt}{\sqrt{t^2-\varepsilon^2}} \frac{e^{3 t} \left(27-e^t \left(e^t \left(8 e^t-31\right)+44\right)\right)}{\left(e^t-1\right)^4 }\times \cosh \left(2 \sqrt{t^2-\varepsilon ^2} \right)~.
%\frac{dt}{\sqrt{t^2-\varepsilon^2}} \frac{   e^{-2 t} (2 \sinh t+29 \cosh t-22)-4}{\left(e^{-t}-1\right)^4 } \\ \times \left(e^{2 t-\sqrt{-s} i \sqrt{t^2-\varepsilon ^2}} + e^{2 t+\sqrt{-s} i \sqrt{t^2-\varepsilon ^2}} \right)
\end{equation}
%We have expressed the result in terms of $s$, because we had to analytically continue one step of the calculation, to ensure convergence of one of the steps. 
Formally setting $\varepsilon = 0$ and taking the small-$t$ expansion of the integrand yields the $\tfrac{1}{t}$ coefficient 
\begin{equation}
\tilde{\mathcal{B}}_{11} = -\frac{599}{120}~,
\end{equation}
associated with the logarithmic divergence. The heat kernel integral $\tilde{\mathcal{I}}_{11}$ is evaluated in appendix \ref{intgrav}, and the resulting expression is provided in (\ref{I11}). 
\newline\newline
For the second set of eigenvalues, after a similar procedure, we find 
%({\color{blue}I must re-check, MISTAKE!!! I summed starting at $n=2$})
\begin{equation}\label{vec2Z}
\tilde{\mathcal{I}}_{12} =  \int_{\varepsilon}^\infty  \frac{dt}{\sqrt{t^2-\varepsilon^2}}\frac{2e^{\frac{t}{2}}  (9 \sinh t+11 \cosh t-5)}{\left(e^t-1\right)^4 } \times  \cosh \left(\tfrac{\sqrt{13}}{2}\sqrt{t^2-\varepsilon ^2}\right)~.
%2 \int_{\varepsilon}^\infty  \frac{dt}{\sqrt{t^2-\varepsilon^2}}\frac{e^{\frac{7 t}{2}} (-9 \sinh t+11 \cosh t-5)}{\left(e^t-1\right)^4 } \times  \cosh \left(\tfrac{\sqrt{13}}{2}\sqrt{t^2-\varepsilon ^2}\right)~.
%{\sqrt{t^2-\varepsilon^2}} \frac{-10 e^{2 t}+5 \sinh t+75 \cosh t-59}{\left(e^t-1\right)^4}  \\ \times \left( e^{\frac{9 t}{2}-i \sqrt{-s} \sqrt{t^2-\varepsilon ^2}} + e^{\frac{9 t}{2}+i \sqrt{-s} \sqrt{t^2-\varepsilon ^2}} \right)~,
\end{equation}
Formally setting $\varepsilon = 0$ and taking the small-$t$ expansion of the integral yields  the $\tfrac{1}{t}$ coefficient 
\begin{equation}
\tilde{\mathcal{B}}_{12} =-\frac{7}{30}~,
\end{equation}
%\begin{equation}
%\frac{s (2 s-9)}{8 t}  = -\frac{65}{64}\frac{1}{t}~,
%\end{equation}
associated to the logarithmic divergence. The heat kernel integral $\tilde{\mathcal{I}}_{12}$ is evaluated in appendix \ref{intgrav}, and the resulting expression is provided in (\ref{I12}).

\subsubsection*{Tensor determinant}

Once again, the relevant eigenvalues are a shifted version of  (\ref{lichspec}),  (\ref{lichspec2}), and  (\ref{lichspec3}). Specifically, we must shift the eigenvalues by $-2\Lambda$. Even after the shift, all eigenvalues remain positive, indicating that the tensor fluctuations of $\mathbb{C}P^2$ are stable. This is similar to the tensor spectrum of $S^4$, but different from that of $S^2\times S^2$ and the Page metric, both of which exhibit one negative mode \cite{Volkov:2000ih,Hennigar:2024gbg}. 
\newline\newline
Taking $\Lambda=\tfrac{3}{2}$, the shifted spectrum is now given by
\begin{eqnarray}\label{lichspec12}
\tilde{\lambda}^{(2)}_n &=& (n+2)(n+4)-3~, \quad\quad \tilde{d}^{(21)}_n = (n+3)^3~,  \quad\quad n \in \mathbb{N}~, \\ \label{lichspec22}
\tilde{\gamma}^{(2)}_n &=& (n+3)(n+4)-3~, \quad\quad \tilde{d}^{(22)}_n = (2n+7)(n+2)(n+5)~, \quad\quad n \in \mathbb{N}~, \\ \label{lichspec32}
\tilde{\delta}^{(2)}_n &=& (n^2+8n+18)-3~, \quad\quad \tilde{d}^{(23)}_n = 2(n+1)(n+4)(n+7)~, \quad\quad n \in \mathbb{N}~. 
\end{eqnarray}
For the first set of tensor modes, namely (\ref{lichspec12}), we find a heat kernel integral that is identical to the first set of vector modes (\ref{vec1Z}), thus
\begin{equation}
\tilde{\mathcal{I}}_{21} = \int_{\varepsilon}^\infty  \frac{dt}{\sqrt{t^2-\varepsilon^2}} \frac{-8 e^{-2 t}+31 e^{-t}+27 e^t-44}{\left(e^t-1\right)^4 }\times \cosh \left(2 \sqrt{t^2-\varepsilon ^2} \right)~.
%\int_{\varepsilon}^\infty  \frac{dt}{\sqrt{t^2-\varepsilon^2}} \frac{e^{3 t} \left(27-e^t \left(e^t \left(8 e^t-31\right)+44\right)\right)}{\left(e^t-1\right)^4 }\times \cosh \left(2 \sqrt{t^2-\varepsilon ^2} \right)~.
%= \int_{\varepsilon}^\infty  \frac{  e^{-2 t} dt}{\sqrt{t^2-\varepsilon^2}} \frac{   (2 \sinh t+29 \cosh t-22)-4 e^{2t}}{\left(e^{-t}-1\right)^4 } \\ \times \left(e^{2 t- i \sqrt{-s}  \sqrt{t^2-\varepsilon ^2}} + e^{2 t+ i \sqrt{-s}  \sqrt{t^2-\varepsilon ^2}} \right)~,
\end{equation}
%where $s=4$. 
Upon formally setting $\varepsilon=0$, we obtain the $\tfrac{1}{t}$ coefficient in the small-$t$ expansion of the integrand to be 
\begin{equation}
\tilde{\mathcal{B}}_{21} = -\frac{599}{120}~.
\end{equation}
The heat kernel integral $\tilde{\mathcal{I}}_{21}$ is evaluated in appendix \ref{intgrav}, and the resulting expression is provided in (\ref{I11}). 
\newline\newline
For the second set of tensor modes, namely (\ref{lichspec22}), we find the following heat kernel integral
\begin{equation}\label{vec22Z}
\tilde{\mathcal{I}}_{22} = \int_{\varepsilon}^\infty  \frac{dt}{\sqrt{t^2-\varepsilon^2}} \frac{-20 e^{-\frac{5 t}{2}}+80 e^{-\frac{3 t}{2}}-118 e^{-\frac{t}{2}}+70 e^{\frac{t}{2}}}{\left(e^t-1\right)^4} \times \cosh \left(\tfrac{ \sqrt{13}}{2} \sqrt{t^2-\varepsilon ^2}\right)~.
%  \frac{ \left(-10 e^{2 t}+5 \sinh t+75 \cosh t-59\right)}{\left(e^t-1\right)^4} \times \\ \left( e^{\frac{9 t}{2}-i \sqrt{-s} \sqrt{t^2-\varepsilon^2}}  + e^{\frac{9 t}{2}+i \sqrt{-s}  \sqrt{t^2-\varepsilon^2}} \right)~,
\end{equation} 
Upon formally setting $\varepsilon=0$, the $\tfrac{1}{t}$ coefficient in the small-$t$ expansion of the integrand is given by
\begin{equation}
\tilde{\mathcal{B}}_{22} =  -\frac{607}{30}~.
% \left(\frac{1}{4} s (2 s-9)-\frac{8737}{480}\right) =
\end{equation}
The heat kernel integral $\tilde{\mathcal{I}}_{22}$ is evaluated in appendix \ref{intgrav}, and the resulting expression is provided in (\ref{I22}). 
%in the small-$t$ expansion of the integrand.
\newline\newline 
For the third set of tensor modes, namely (\ref{lichspec32}), we find the following heat kernel integral
\begin{equation}\label{vec23Z}
\tilde{\mathcal{I}}_{23} = 
%\int_{\varepsilon}^\infty  \frac{dt}{\sqrt{t^2-\varepsilon^2}} \frac{-20 e^{-\frac{5 t}{2}}+80 e^{-\frac{3 t}{2}}-118 e^{-\frac{t}{2}}+70 e^{t/2}}\times \cosh \left( \sqrt{t^2-\varepsilon ^2}\right) ~.
 \int_{\varepsilon}^\infty  \frac{dt}{\sqrt{t^2-\varepsilon^2}} \frac{4 e^{-t}\left(9 \sinh t+19 \cosh t-16 \right) }{\left(e^t-1\right)^4}\times \cosh \left( \sqrt{t^2-\varepsilon ^2}\right) ~.
%\frac{4 e^{5 t} (-9 \sinh (t)+19 \cosh (t)-16) \cosh \left(\sqrt{s} \sqrt{t^2-\epsilon ^2}\right)}{\left(e^t-1\right)^4 }~.
%\frac{e^t \left(56 e^{3 t} \cosh \left(\sqrt{s} \sqrt{t^2-\epsilon ^2}\right)+5 e^t-16\right)}{\left(e^t-1\right)^4}~.
%\frac{2 \left(e^t \left(5 e^t-16\right)+14\right) }{\left(e^t-1\right)^4} \times \left( e^{4 t+ i \sqrt{-s}  \sqrt{t^2-\varepsilon^2}}+e^{4 t- i \sqrt{-s}  \sqrt{t^2-\varepsilon^2}}\right)~,
\end{equation} 
Upon formally setting $\varepsilon=0$, the $\tfrac{1}{t}$ coefficient in the small-$t$ expansion of the integrand is given by
\begin{equation}
%\frac{30 s^2-540 s+2251}{120}\frac{1}{t} = 
\tilde{\mathcal{B}}_{23} =\frac{1741}{60}~.
\end{equation}
The heat kernel integral $\tilde{\mathcal{I}}_{23}$ is evaluated in appendix \ref{intgrav}, and the resulting expression is provided in (\ref{I23}). 
\newline\newline
%in the small-$t$ expansion of the integrand. 
%\newline\newline
We have now assembled all the necessary pieces to compute the full one-loop gravitational contribution (\ref{Zgrav1}). Combining the results from appendix \ref{intgrav}, we find that the functional determinant ratio in (\ref{Zgrav1}) evaluates to the following heat kernel regularised expression
\begin{equation}
\log {\frac{{{\det}'^{\tfrac{1}{2}} \left( \frac{\Delta_{(1)} - 2\Lambda}{\Lambda_{\text{u.v.}}}\right)}}{{{\det}^{\tfrac{1}{2}} \left( \frac{\Delta_{(2)} - 2\Lambda}{\Lambda_{\text{u.v.}}} \right)}}} = \frac{8}{\varepsilon^4}-\frac{16}{\varepsilon ^2}+ 12 \log A-\frac{541}{60} \log  e^{\gamma } \varepsilon  +2 \zeta '(-3)+\frac{22}{3}+\frac{61}{60} \log 2~.
\end{equation}
The combined result is significantly simpler than each contributing part.  Altogether,
% ({\color{blue}fix}) 
\begin{multline}\label{ZCP2f}
\log | \mathcal{Z}^{(1)}_{\mathbb{C}P^2} | = \frac{18}{\varepsilon^4 \Lambda^2}-\frac{24}{\varepsilon^2 \Lambda}  -\frac{359}{60} \log  \left( \sqrt{\frac{3}{\Lambda}} \frac{2}{ \varepsilon \, e^{\gamma }} \right)  + {\frac{25}{3} } -\frac{1081}{120}  \log 2 + \frac{7}{2}\log 3 \\  -12\zeta'(-1) +2 \zeta '(-3)   - 4 \log \frac{\mathcal{S}_{\mathbb{C}P^2}}{2\pi}  - \log \text{vol}_{\text{can}}~,
% \frac{18}{\varepsilon^4 \Lambda^2}-\frac{24}{\varepsilon^2 \Lambda}  + \log \mathcal{A}+\frac{359}{60} \log  \left( \sqrt{\frac{\Lambda}{3}} e^{\gamma } \varepsilon \right)  +\frac{22}{3}  -\frac{2279}{120}  \log 2 \\  -\frac{15}{2} \log \pi + 12 \log A +2 \zeta '(-3)   - 4 \log {\mathcal{S}_{\mathbb{C}P^2}}  - \log \text{vol}_{\text{can}}~,
\end{multline}
where $\text{vol}_{\text{can}} = \tfrac{\sqrt{3}\pi^5}{{3}}$, we have reinstated $\Lambda$, and used that  the Glaisher constant satisfies $\log A= \tfrac{1}{12}-\zeta'(-1)$. Up to a sign, the overall phase of $ \mathcal{Z}^{(1)}_{\mathbb{C}P^2}$ is $i$. 
% $\mathcal{A}$ was given in (\ref{AZ}), 

\subsection{Comparing to the heat kernel coefficient}

As for the scalar field, the logarithmic coefficient can also be calculated through other means, and the general result for an Einstein metric satisfying $R_{\mu\nu} = \Lambda g_{\mu\nu}$ follows from equation (4.15) of \cite{Christensen:1979iy} which for linearised general relativity with $\Lambda>0$ yields
\begin{equation}\label{B4t}
B^{{\mathbb{C}P^2}}_4 =  \frac{1}{180(4\pi)^2}\left(212 \times 32\pi^2 \chi_{\mathbb{C}P^2} -2088 \Lambda^2 V_{\mathbb{C}P^2}   \right) = -\frac{359}{60}~.
\end{equation}
If we sum the coefficients of the various $\tfrac{1}{t}$ coefficients arising in the path integral computation we find 
\begin{equation}
\left(-\frac{599}{120}-\frac{607}{30}+\frac{1741}{60}\right)-\left(-\frac{599}{120}-\frac{7}{30}\right) - 2 \times 8 + 1 = \frac{541}{60}- 15 = -\frac{359}{60}~.
%-\frac{659}{120}~.
%- \frac{89}{120} + \frac{19}{15} - \left( \frac{31}{120} -1 \right) + 2 = -\frac{11}{15}~.
\end{equation}
We note that, as is indicated in the $\Lambda_{\text{u.v.}}$ dependent  pre-factor of (\ref{Zgrav1}), the final $-2\times 8+1$ stems from a proper treatment of the path integration measure. A similar treatment can be done for the linearised Einstein theory about the round $S^4$ saddle, see for example  \cite{Volkov:2000ih,Anninos:2020hfj}, which yields a coefficient 
\begin{equation}
B^{S^4}_4 = \frac{89}{18}  - \left( -\frac{191}{30} \right ) - (10+5)\times 2 + (5+1) = -\frac{571}{45}~,
\end{equation}
where we have expressed the coefficient in terms of the transverse traceless tensor, transverse  vector, and zero mode split. This indeed agrees with the $S^4$ version of (\ref{B4t}). The full heat kernel regularised expression for $S^4$, given in (C.46) of \cite{Anninos:2020hfj}, reads
\begin{multline}\label{ZS41}
\log |\mathcal{Z}^{(1)}_{S^4}| = \frac{24}{\varepsilon^4 \Lambda^2} - \frac{32}{\varepsilon^2 \Lambda}-\frac{571}{45}\log \left( \sqrt{\frac{3}{\Lambda}} \frac{2 }{\varepsilon \, e^{\gamma} }   \right)+\frac{715}{48} -\log 2  \\ -\frac{47}{3}\zeta'(-1)+\frac{2}{3}\zeta'(-3)-5 \log \frac{\mathcal{S}_{S^4}}{2\pi}  - \log \text{vol}_{\text{can}} ~,
\end{multline}
where now $\text{vol}_{\text{can}} = \tfrac{2}{3}(2\pi)^6$ is the canonically normalised group volume of the $SO(5)$ isometry group of $S^4$. The overall phase of $\mathcal{Z}^{(1)}_{S^4}$ is $(\pm i)^{1+5} = -1$ \cite{Polchinski:1988ua}. 
\newline\newline
We can compare the two leading heat kernel coefficients of the four-sphere one-loop expression (\ref{ZS41}) to the corresponding ones (\ref{ZCP2f}) for $\mathbb{C}P^2$, and find that their ratio is indeed $\tfrac{4}{3}$. Our heat kernel coefficients for $\mathbb{C}P^2$ also agree with those computed in \cite{Vassilevich:1993yt} up to zero mode omissions. 
%to compare take $Lambda=3/2$. for maxwell -R/3 = -2, and for gravity -4 R/3 = -8 
Interestingly,  the finite parts of the two expressions (\ref{ZCP2f}) and (\ref{ZS41}) have the same degree of transcendentality. 
\newline\newline
For the sake of completeness we also mention that the heat kernel coefficient of $S^2 \times S^2$ as computed by (\ref{B4t}) is $B^{{S^2\times S^2}}_4 = -\tfrac{98}{45}$, in agreement with the functional determinant computation of \cite{Volkov:2000ih}. For the Page metric on $\mathbb{C}P^2 \# \overline{\mathbb{C}P^2}$ one finds an irrational coefficient, approximately\footnote{More precisely, using the exact expressions in \cite{Page:1978vqj}, one finds that the heat kernel coefficient is
\begin{equation}
B^{\mathbb{C}P^2 \# \overline{\mathbb{C}P^2}}_4 = \frac{\bold{x}}{90}
\end{equation}
where $\bold{x}$ is the largest real root of $x^4+2872 x^3+1049514 x^2+821718616 x+108404462761 = 0~.$
%\begin{equation}
%
%\end{equation}
}
$B^{\mathbb{C}P^2 \# \overline{\mathbb{C}P^2}}_4 = -1.66$. Finally, $B^{\mathbb{C}P^2 \# k \overline{\mathbb{C}P^2}}_4 = -\tfrac{359}{60} +\tfrac{137}{36} \tiny{k}$ for $k=3,\ldots,8$.
\newline\newline
More generally, it will be interesting to understand the  complete one-loop (and all-loop \cite{Bandaru:2024qvv,Muhlmann:2022duj,Anninos:2024fty}) structure of the partition function of the Universe (\ref{ZU}) for all known Einstein metrics with positive scalar curvature. Perhaps this dataset is sufficiently large so as to encode many of the physical properties of the low energy effective field theory and beyond.

\section*{Acknowledgements}

It is a great pleasure to acknowledge discussions with Alejandra Castro, Dami\'an Galante, Pietro Benetti Genolini, Thomas Hertog, Chris Herzog, Albert Law, Juan Maldacena, Miguel Montero,  Beatrix M\"uhlmann, and Edgar Shaghoulian. We are very grateful to Themistoklis Zikopoulos for pointing out the volume of $V_{\mathbb{C}P^2 \# 2 \overline{\mathbb{C}P^2}}$. D.A. is funded by the Royal
Society under the grant ``Concrete Calculables in Quantum de Sitter" and the STFC Consolidated grant ST/X000753/1. C.B. is funded by STFC under the grant reference STFC/2887726. S.B. is funded by STFC under the grant reference STFC/2928477. 

\appendix

\section{A  test of the scalar partition function}\label{scalarZapp}

In this appendix, we present a test of the free massive scalar path integral on $\mathbb{C}P^2$. It will be useful to  recall that the Fubini-Study metric of $\mathbb{C}P^2$ can be obtained from an embedding space picture. We follow the description in \cite{Pope:1980ub}. Consider an $S^5$ embedded in $\mathbb{C}^3$, endowed with complex coordinates $\bold{Z} \equiv (Z^1,Z^2,Z^3)$
\begin{equation}
|Z^1|^2 + |Z^2|^2  + |Z^3|^2  = \frac{6}{\Lambda}~.
\end{equation}
When considering the embedding space picture it is convenient to set $\Lambda=6$, and we will do so in subsection \ref{appA2}. 
\newline\newline
To obtain $\mathbb{C}P^2$ we further identify the $U(1)$ action $Z^I \to e^{i\alpha} Z^I$. If we define coordinates $\zeta^1 = \frac{Z^1}{Z^3}$, $\zeta^2 = \frac{Z^2}{Z^3}$, and $e^{i\alpha} = \frac{Z^3}{|Z^3|}$, the induced metric on $S^5$ is given by an $S^1$ fibered over $\mathbb{C}P^2$, where $\alpha \in (0,2\pi]$ denotes the fiber coordinate. We can thus understand the Laplacian on $S^5$ in terms of that of a charged particle on $\mathbb{C}P^2$ in the presence of a background $U(1)$ gauge field given by
\begin{equation}
A = \frac{i}{2} \left(1 + \zeta^1  \bar{\zeta}^1 + \zeta^2  \bar{\zeta}^2  \right)^{-1} \left( \bar{\zeta}^1 d\zeta^1 + \bar{\zeta}^2 d\zeta^2 -   d\bar{\zeta}^1 \zeta^1 - d\bar{\zeta}^2 \zeta^2 \right)~,
\end{equation}
 where  we have expressed the result in units $\Lambda=6$. The charge of the particle is given by the momentum, $e \in \mathbb{Z}$, along the $S^1$. The geodesic distance $D=\cos^{-1} \sigma$ between two points, $\zeta^i$ and $\xi^i$, on $\mathbb{C}P^2$ (with $\Lambda=6$) is given by
\begin{equation}
\sigma = \frac{|1 + \delta_{i\bar{j}} \zeta^{i}  \bar{\xi}^{\bar{j}}|}{\sqrt{(1+ \delta_{i\bar{j}}\zeta^i \bar{\zeta}^{\bar{j}})(1+ \delta_{i\bar{j}} \xi^i \bar{\xi}^{\bar{j}})}}~.
%$\rho = 1+\zeta^i \bar{\zeta}^i$
 \end{equation}
At coincidence $\sigma=1$.

\subsection{Comparison to coincident point limit of Green's function}

We would like to consider the Green's function of a scalar field of mass squared $\mu^2$ on $\mathbb{C}P^2$. This exercise was undertaken in \cite{Warner:1982fv}, where it is shown that the scalar Green's function is given by
\begin{equation}
G_{\mathbb{C}P^2}(\sigma) = \frac{\Gamma (\Delta ) \Gamma (\bar{\Delta}) }  {16\pi^2   \Gamma (1+|e|) }\sigma^{\frac{|e|}{2}} {_2 F}_1 \left(\Delta,\bar{\Delta}, 1+|e|, \sigma \right)~,
\end{equation}
%\Gamma (|e|+2-\Delta -\bar{\Delta})\csc (\pi  (|e|-\Delta -\bar{\Delta}+1)
%The geodesic distance $\sigma$ represents the distance between two points in $\mathbb{C}P^2$, such that the coincident point is at $\sigma=1$. 
with
\begin{equation}
\Delta \equiv 1 + \frac{|e|}{2} + \frac{i}{2} \sqrt{\frac{6}{\Lambda} \mu^2-4-e^2}~, \quad \bar{\Delta} \equiv 1 + \frac{|e|}{2} - \frac{i}{2} \sqrt{\frac{6}{\Lambda} \mu^2-4-e^2}~.
\end{equation}
To compare with the main text, we will be interested in the neutral sector $e=0$, which we take in what follows. In this subsection we take $\Lambda=\tfrac{3}{2}$ from now on. The Green's function $G_{\mathbb{C}P^2}(\sigma)$ obeys a standard coincident point normalisation
\begin{equation}
G_{\mathbb{C}P^2}(\sigma) \approx \frac{1}{16\pi^2 (1-\sigma)}~. 
\end{equation}
Although the coincident point limit of $G_{\mathbb{C}P^2}(\sigma)$ is divergent, upon taking sufficient derivatives with respect to $\mu^2$, we obtain a finite expression. This allows us to compare to the appropriate number of $\mu^2$ derivatives of (\ref{logZphi}) in the main text. This is because a $\mu^2$ derivative of $\log \mathcal{Z}^{(\varphi)}_{\mathbb{C}P^2}$ computes the coincident point limit of $G_{\mathbb{C}P^2}(\sigma)$ integrated over all of $\mathbb{C}P^2$ times minus one-half. Let us test this reasoning by computing the fourth $\mu^2$ derivative of $\log \mathcal{Z}^{(\varphi)}_{\mathbb{C}P^2}$, for the specific value of $\mu^2 = 1$. From expression (\ref{logZphi}), we find
\begin{equation}\label{Zpppp}
\partial_{\mu^2}^{(4)} \log \mathcal{Z}^{(\varphi)}_{\mathbb{C}P^2}|_{\mu^2=1} = 3\zeta(5)~.
\end{equation}
We note that we can set $\varepsilon=0$, and perform the $t$-integral, since the expression is ultraviolet finite. We  now compute 
\begin{equation}
\lim_{\sigma\to 1} \partial_{\mu^2}^{(3)} G_{\mathbb{C}P^2}(\sigma)|_{\mu^2=1} = - \frac{3 \zeta(5)}{4\pi^2}~.
\end{equation}
To do so, we have used standard series expansions in Mathematica. Recalling that the volume of $\mathbb{C}P^2$ in units where $\Lambda=\tfrac{3}{2}$ is $8\pi^2$, we have that
\begin{equation}
-\frac{1}{2}  \times V_{\mathbb{C}P^2} \times \lim_{\sigma\to 1} \partial_{\mu^2}^{(3)} G_{\mathbb{C}P^2}(\sigma)|_{\mu^2=1}  = \partial_{\mu^2}^{(4)} \log \mathcal{Z}^{(\varphi)}_{\mathbb{C}P^2}|_{\mu^2=1}~.
\end{equation}
This tests our expression against the Green's function which does not explicitly use the spectrum of the scalar Laplacian in its construction.

\subsection{Integral representation of Green's function}\label{appA2}
%https://functions.wolfram.com/HypergeometricFunctions/Hypergeometric2F1/07/01/01/
We can also proceed from an embedding space perspective, setting $\Lambda=6$. First, we recall that on an $S^5$, the Green's function can be represented as (see (2.36) of \cite{Bandaru:2024qvv})
\begin{equation}
G_{S^5}({X},{Y}) = \frac{\Gamma\left(\frac{\boldsymbol{\Delta}+\bar{\boldsymbol{\Delta}}}{2}\right)}{4\pi^{3}}  \int_0^1 \frac{d\lambda}{\lambda} \frac{\lambda^{\boldsymbol{\Delta}} + \lambda^{\bar{\boldsymbol{\Delta}}}}{\left(1+\lambda^2-2\lambda\sigma\right)^{\frac{\Delta+\bar{\Delta}}{2}}}~.
%\int_\lambda^* \frac{\lambda^\Delta}{\left(1+\lambda^2-2\lambda\sigma\right)^{\frac{\Delta+\bar{\Delta}}{2}}}~.
\end{equation}
In the above, ${X}$ and ${Y}$ are unit normal vectors on $\mathbb{R}^6$, $\sigma \equiv {X} \cdot {Y}$, and now we have
\begin{equation}
\boldsymbol{\Delta} = 2 + i \sqrt{\mu^2 - 4}~,  \quad \bar{\boldsymbol{\Delta}} = 2 - i \sqrt{\mu^2 - 4}~.
\end{equation}
To make contact with $\mathbb{C}P^2$ we can collect the coordinates on $\mathbb{R}^6$ into complex pairs, such that 
\begin{eqnarray}
Z^1 &=& X^1 + i X^2~,  \quad Z^2 = X^3 + i X^4~, \quad Z^3 = X^5 + i X^6~, \\  
W^1 &=& Y^1 + i Y^2~,  \quad W^2 = Y^3 + i Y^4~, \quad W^3 = Y^5 +  i Y^6~. 
\end{eqnarray}
As already mentioned, the coordinates on $\mathbb{C}P^2$ can be written as  $\zeta^1 = \frac{Z^1}{Z^3}$, $\zeta^2 = \frac{Z^2}{Z^3}$, and $e^{i\alpha} = \frac{Z^3}{|Z^3|}$, and similarly $\upsilon^1 = \frac{W^1}{W^3}$, $\upsilon^2 = \frac{W^2}{W^3}$, and $e^{i\beta} = \frac{W^3}{|W^3|}$. We can thus extract the neutral $\mathbb{C}P^2$ Green's function by integrating $G_{S^5}({X},{Y})$ over $\alpha \in (0,2\pi]$. (To do so, it is convenient to place one of the two points at the origin of $\mathbb{C}P^2$.) This yields the coincident point formal expression
\begin{equation}\label{GCP2}
G_{\mathbb{C}P^2}(\zeta^i,\zeta^i) = \frac{1}{4\pi^3} \times 2\pi \int_0^1 \frac{d\lambda}{2\lambda} (1+\lambda) \frac{\lambda^{\frac{\boldsymbol{\Delta}}{2}} + \lambda^{\frac{\bar{\boldsymbol{\Delta}}}{2}}}{(1-\lambda)^3}~. 
\end{equation}
Once again, we can compute
\begin{equation}
-\frac{1}{2}  \times V_{\mathbb{C}P^2} \times \lim_{\sigma\to 1} \partial_{\mu^2}^{(3)} G_{\mathbb{C}P^2}(\sigma)|_{\mu^2=4}  = \frac{3 \zeta (5)}{256}~, 
\end{equation}
where we recall that now we are in units where $\Lambda = 6$, such that the volume of $\mathbb{C}P^2$ is $\tfrac{\pi^2}{2}$. This change in units accounts for the relative factor of $256 = 4^4$ in (\ref{Zpppp}), and thus again we have agreement. 
\newline\newline
It is interesting that the expression (\ref{GCP2}) takes a character-like form resembling the expressions in \cite{Anninos:2020hfj} for the sphere. To do so it is convenient to use that $- \lambda \partial_\lambda \partial_\nu \tfrac{\lambda^{\pm i \nu}}{\log\lambda} = \nu \lambda^{\pm i \nu}$. Denoting ${\boldsymbol{\Delta}}= 2 + 2 i \nu$, with $\nu = \sqrt{\tfrac{1}{4} \mu^2 - 1}$, and upon integrating by parts in $\lambda$, we can re-express  (\ref{GCP2}) as
\begin{equation}\label{frederik}
-\frac{1}{2} \times V_{\mathbb{C}P^2} \times G_{\mathbb{C}P^2}(\zeta^i,\zeta^i) =   \frac{1}{8 \nu} \partial_\nu \int_0^1 \frac{d\lambda}{2 \lambda (-\log \lambda)} \frac{1+4\lambda+\lambda^2}{1-\lambda} \frac{\lambda^{\frac{\boldsymbol{\Delta}}{2}} + \lambda^{\frac{\bar{\boldsymbol{\Delta}}}{2}}}{(1-\lambda)^3}~.
%\frac{\pi^2}{2} 
\end{equation}
Noting that $\partial_\nu = 8 \nu \partial_{\mu^2}$, we are led to expression (\ref{logZphi}) where $\varepsilon$ has been formally set to zero.

\section{Explicit expressions for heat kernel integrals}\label{integrals}

In this appendix, we present explicit expressions for the various heat kernel integrals that appear in the main text. The method we use is described in detail in appendix C of \cite{Anninos:2020hfj}. The essence of the method is to split the heat kernel integral into an ultraviolet part and infrared part that. The ultraviolet part can be efficiently computed in terms of the small-$t$ expansion of the integrand of the heat kernel integral (see (C.12) of \cite{Anninos:2020hfj}). The infrared part can be efficiently computed in terms of special functions (see (C.19) of \cite{Anninos:2020hfj}), and in particular the Hurwitz zeta function defined as
\begin{equation}
\zeta(z,a) \equiv \sum_{n=0}^\infty (n+a)^{-z}~,
\end{equation}
which admits an analytic continuation to a meromorphic function in the whole complex-$z$ plane. The Riemann zeta function $\zeta(z) = \zeta(z,1)$. We have tested our results against numerical evaluation of the heat kernel integrals for values up to $\varepsilon \sim 10^{-2}$ and the comparison falls within the $\mathcal{O}(\varepsilon)$ error. 

\subsection{Maxwell gauge field}\label{intmax}

The heat kernel integral (\ref{massless}) evaluates to
\begin{equation}\label{I0}
\mathcal{I}_0 = \frac{4}{\varepsilon ^4}+\frac{1}{\varepsilon ^2} -3 \log A+\frac{89}{120} \log  e^{\gamma } \varepsilon +\zeta '(-3)+-\frac{1}{12}-\frac{29}{120}  \log 2~,
\end{equation}
where $A \approx 1.282$ denotes the Glaisher constant, and $\gamma$ is the Euler-Mascheroni constant  here and below. This is also the expression for $\mathcal{I}_{11}$ in (\ref{vector1}), as it is the same integral. 
\newline\newline
The heat kernel integral (\ref{vector2}) evaluates to
\begin{equation}\label{I1}
\mathcal{I}_{12} = \frac{8}{\varepsilon ^4}-\frac{4}{\varepsilon ^2}-\frac{19}{15}  \log \frac{\varepsilon e^\gamma }{2} + 3 \log A+2 \zeta'(-3) +\frac{13}{48}~.
\end{equation}
It is worth recalling that the Glaisher constant obeys the identity $\log A = \tfrac{1}{12} - \zeta'(-1)$.

\subsection{Linearised graviton}\label{intgrav}

The heat kernel integral (\ref{vec1Z}) evaluates to
\begin{equation}\label{I11}
\tilde{\mathcal{I}}_{11} =  \frac{4}{\varepsilon ^4}+ \frac{4}{\varepsilon^2} -12 \log A+\frac{599}{120} \log  \frac{e^{\gamma } \varepsilon }{2} +\zeta '(-3)+-\frac{13}{3}+\frac{\log 3}{2}+8 \log 2~.
\end{equation}
The heat kernel integral (\ref{vec2Z}) evaluates to
\begin{equation}\label{I12}
\tilde{\mathcal{I}}_{12} = \frac{8}{\varepsilon^4}+\frac{2}{\varepsilon^2}+ \frac{13}{48}+\frac{7}{30} \log \frac{\varepsilon e^\gamma}{2}  + 
\frac{1}{2} \sum_{\pm} P(\hat{\delta}-\Delta_\pm) \zeta'(0,\Delta_\pm)~,
\end{equation}
where $\Delta_\pm = \tfrac{5}{2} \pm  \tfrac{\sqrt{13}}{2}$ and $P_{12}(n) \equiv (n+1) (n+4) (2 n+5)$.
\newline\newline
The heat kernel integral (\ref{vec22Z}) evaluates to
\begin{equation}\label{I22}
\tilde{\mathcal{I}}_{22} = \frac{8}{\varepsilon ^4}+\frac{2}{\varepsilon ^2}+\frac{13}{48} + \frac{607}{30} \log \frac{e^{\gamma } \varepsilon }{2} + \frac{1}{2} \sum_{\pm} P(\hat{\delta}-\Delta_\pm) \zeta'(0,\Delta_\pm)~,
\end{equation}
where $\Delta_\pm = \tfrac{7}{2} \pm  \tfrac{\sqrt{13}}{2}$ and $P_{22}(n) \equiv (n+2) (n+5) (2 n+7)$. We note that
\begin{equation}
\tilde{\mathcal{I}}_{22}  - \tilde{\mathcal{I}}_{21} =  20 \log  \frac{\sqrt{3}e^{\gamma } \varepsilon }{2} = \frac{1}{2} \times 20 \log \frac{2\Lambda }{ \Lambda_{\text{u.v.}}}~.
\end{equation}
The above also follows directly by noting that  the spectra (\ref{lichspec22}) and (\ref{spec12}) agree up to a single eigenvalue $\tilde{\gamma}_1^{(1)} =2\Lambda$ of degeneracy $20$.
\newline\newline
The heat kernel integral (\ref{vec23Z}) evaluates to
\begin{equation}\label{I23}
\tilde{\mathcal{I}}_{23} = \frac{8}{\varepsilon ^4}-\frac{16}{\varepsilon ^2}-\frac{1741   }{60}\log e^\gamma \varepsilon + 12 \log A+2 \zeta '(-3)+\frac{22}{3}+\frac{1261 }{60}\log 2-10 \log 3~.
\end{equation}
%where $\Delta_\pm = 4 \pm 1$ and $P_{23}(n) \equiv $.

\section{Comments on $S^2 \times S^2$ saddle}\label{S2S2app}

In this appendix we review the spectrum of $S^2 \times S^2$ Einsten metric, as calculated in \cite{Volkov:2000ih}, and some properties of the corresponding one-loop correction.

\subsection{Vector spectrum}

The spectrum of the vector operator $\Delta_{(1)}-2\Lambda$ on $S^2\times S^2$ is given by
\begin{equation}
\tilde{\lambda}^{(1)}_n = \left( n(n+1)-2 \right)\Lambda~, \quad\quad \tilde{d}_n^{(11)}  =  2(2n+1)~,
\end{equation}
with $n  \in \mathbb{Z}^+ + 1$. In addition we have the eigenmodes
\begin{equation}\label{v2spec}
\tilde{\gamma}^{(1)}_n = \left(n_1(n_1+1)+ n_2(n_2+1)-2\right)\Lambda~, \quad\quad \tilde{d}_{n_1 n_2}^{(12)} = 3(2n_1+1)(2n_2+1)~, 
\end{equation}
with $n_1, n_2 \in \mathbb{Z}^+$. 

\subsection{Tensor spectrum}

The transverse-traceless spectrum of the operator $\Delta_{(2)}-2\Lambda$ on  $S^2\times S^2$ is given by
\begin{equation}
\tilde{\lambda}^{(2)}_n = (n(n+1)-2)\Lambda~, \quad\quad \tilde{d}_n^{(21)}  =  2(2n+1)~, 
\end{equation}
with $n  \in \mathbb{Z}^++1$. Moreover, we have
\begin{equation}
\tilde{\gamma}^{(2)}_n = n(n+1)\Lambda~, \quad\quad \tilde{d}_n^{(22)}  =  18(2n+1)~, 
\end{equation}
with $n   \in \mathbb{Z}^+ +1$, and finally 
\begin{equation}\label{t3spec}
\tilde{\delta}^{(2)}_n = \left(n_1(n_1+1)+ n_2(n_2+1)-2\right)\Lambda~, \quad\quad \tilde{d}_{n_1 n_2}^{(23)} = 5(2n_1+1)(2n_2+1)~, 
\end{equation}
with $n_1,n_2  \in \mathbb{Z}^+ +1$. In addition there is a single negative tensor eigenmode with eigenvalue $-2\Lambda$, and nine tensor eigenmodes with eigenvalue $2\Lambda$. The negative mode corrresponds to changing the relative size of the two-spheres without breaking any spherical symmetry.  

\subsection{One-loop contribution}

The one-loop determinants stemming from the spectrum (\ref{v2spec}) and (\ref{t3spec}) of $S^2\times S^2$ are captured by the general heat kernel regularised formula
\begin{equation}
\mathcal{I}_{a,\mu} = \sum_{n_1, n_2 \ge a} (2n_1+1) (2n_2+1) \int_{\mathbb{R}^+} \frac{d\tau}{2\tau} e^{- \frac{\varepsilon^2}{4\tau}}e^{-\left( n_1(n_1+1) \Lambda + \mu^2\right)\tau}e^{-\left( n_2(n_2+1) \Lambda + \mu^2\right) \tau}~.
\end{equation}
For the spectrum (\ref{t3spec}), we have $\mu^2 = -\Lambda$ and $a=2$. Once again, this can be treated using the method in appendix C of \cite{Anninos:2020hfj}. The essential difference is that now the spectrum is labelled by two different integers. One way to view this as an infinite Kaluza-Klein type tower of particles in dS$_2$ with some low lying modes removed. Upon rescaling $t \rightarrow \tfrac{t}{\sqrt{\Lambda}}$, $\varepsilon \rightarrow \tfrac{\varepsilon}{\sqrt{\Lambda}}$, we  obtain the following sum over heat kernel integrals
%The essential difference is that now we introduce two different $u$-integrals. 
%One can use polar coordinates for the $(u_1,u_2)$-plane to integrate out the angle, yielding the following integral
\begin{multline}\label{S2S21}
\mathcal{I}_{a,\mu} = \sum_{n_1 \ge a} (2n_1+1) \int_{\varepsilon}^\infty \frac{dt}{2\sqrt{t^2-\varepsilon^2}} \frac{e^{-(a+1) t} \left(e^t+1+2 a \left(e^t-1\right)\right)}{\left(1-e^{-t}\right)}  \\ \times  \frac{e^{-\frac{t}{2}-i \nu_{n_1} \sqrt{t^2-\varepsilon^2}}+e^{-\frac{t}{2}+i \nu_{n_1} \sqrt{t^2-\varepsilon^2}}}{\left( 1-e^{-t}\right)}~, %+  e^{i \nu_{n_1} \sqrt{t^2-\varepsilon^2}
%\mathcal{I}_{a,\mu} = \sqrt{2\pi} \int_{\mathbb{R}^+} \frac{d\rho}{\sqrt{\rho^2+\varepsilon^2}} K_1\left(\sqrt{2}\nu \rho \right)  J_0\left(\rho\sqrt{\left(n_1+\frac{1}{2}\right)^2+\left(n_2+\frac{1}{2}\right)^2}  \right)~,
%\int_\varepsilon^\infty \frac{dt}{2\sqrt{t^2-\varepsilon^2}} \frac{2 e^{-t} +a \left(1-e^{-t}\right)}{ \left(1-e^{-t}\right)^2 } \left( e^{-\frac{t}{2}-i \nu  \sqrt{t^2-\varepsilon ^2}}+e^{-\frac{t}{2}+i \nu  \sqrt{t^2-\varepsilon ^2}} \right)~,
\end{multline}
where we have defined the Kaluza-Klein mass spectrum
\begin{equation}
 \nu_{n_1}^2 \equiv n_1\left(n_1+1\right)   +2 {\mu^2}{\Lambda^{-1}} - \frac{1}{4} ~.
 \end{equation}
%with $\nu^2 \equiv \tfrac{\mu^2}{\Lambda}  - \tfrac{1}{4}$. A straightforward application of this methods indicates that the resulting expression will involve terms of the type $\zeta'(-n,\frac{1}{2} \pm \sqrt{5})$ with $n=0,1$.
For the case at hand, $\mu^2\Lambda^{-1} = -1$, such that the $\nu_{n_1}$ are irrational. Unlike what happens for $S^4$ and $\mathbb{C}P^2$, the finite piece cannot be reduced to a simple expression in terms of derivatives of the Riemann zeta function. On the other hand, since we have an infinite sum over $SO(2,1)$ characters in (\ref{S2S21}) furnishing the principal series irreducible representation with $\Delta_{n_1} =\tfrac{1}{2}+i \nu_{n_1}$, perhaps we can view these as enhanced (possibly non-unitary) contributions to the edge mode sector of (\ref{ZdS}), itself built from $SO(2,1)$ characters (see also \cite{Law:2025yec,Mukherjee:2025xlt}).

%If we formally set $\epsilon=0$, we find that the $\tfrac{1}{t}$ coefficient of the small-$t$ expansion of the integrand  is equal to $-66$.
%\newline\newline
%The heat kernel integral evaluates to
%\begin{equation}
%\mathcal{I}^{(22)} = 18 \zeta '(-1,3)-9 \zeta'(0,3)+18\zeta'(-1,2)+9 \zeta'(0,2)-\frac{9}{2}-66 \log \frac{2 e^{-\gamma}}{\epsilon} +\frac{36}{\epsilon^2} ~.
%\end{equation}

\bibliographystyle{JHEP}
\bibliography{bib}
\end{document}